\documentclass[11pt,a4paper]{article}
\usepackage{jheppub}
\usepackage{booktabs}
\usepackage{tikz}
\usetikzlibrary{decorations.pathmorphing,decorations.markings,arrows.meta,calc,positioning}
\newcommand{\dd}{\mathrm{d}}
\usepackage[english]{babel}
\usepackage{microtype}
\usepackage{tcolorbox}
\usepackage[normalem]{ulem}
\definecolor{lblue}{RGB}{45,140,235}

\newcommand{\radd}[1]{{\color{lblue}#1}}

\preprint{MPP-2026-138}

\newcommand{\Om}{\Omega}
\newcommand{\vk}{\vec k}
\newcommand{\Tri}{T}

\newcommand{\cS}{\mathcal S}
\newcommand{\ox}{\otimes}
\newcommand{\pT}{\widetilde{T}}

\title{A compact analytic formula for the one-loop triangle cosmological correlator}

\author[a]{Johannes M.\ Henn,}
\author[b]{Jiajie Mei,}
\author[a]{and Qinglin Yang}

\affiliation[a]{Max-Planck-Institut f\"ur Physik, Werner-Heisenberg-Institut,
Boltzmannstra{\ss}e 8, 85748 Garching, Germany}
\affiliation[b]{Institute for Theoretical Physics, University of Amsterdam,
Science Park 904, 1098 XH Amsterdam, The Netherlands}

\emailAdd{henn@mpp.mpg.de}
\emailAdd{j.mei@uva.nl}
\emailAdd{qlyang@mpp.mpg.de}

\abstract{We derive a compact analytic formula for the one-loop triangle correlator
of conformally coupled scalars in de Sitter space. The result is organised as six
leading-singularity prefactors multiplying pure weight-two functions of the six energy
variables. It contains forty-two dilogarithms, compared with approximately one hundred
and twenty in the previously known closed-form representation, and requires no auxiliary
regulator. The dilogarithms occur in Galois-conjugate pairs, making each contribution
separately real throughout the physical region. We validate the result numerically. Moreover, we
show that its symbol can be derived directly from the dressed integral representation or,
independently, bootstrapped from Landau singularities and general consistency conditions.
Finally, we show that the correlator (in a suitable normalization) is a
Stieltjes function of each squared energy separately and is jointly completely monotone
in all six squared energies. These structures suggest a route towards higher-point
one-loop cosmological correlators.
}

\begin{document}
\maketitle
\flushbottom

\newpage

\section{Introduction}
\label{sec:intro}

Cosmological correlators encode the statistics of primordial fluctuations at late times;
see~\cite{Baumann:2022jpr} for a review. They provide the theoretical quantities from
which observations of the cosmic microwave background and the distribution of galaxies
are inferred. In particular, their non-Gaussian components retain information about the
interactions and particle content present during inflation. Understanding the analytic
structure of these correlators is therefore important both for extracting this information
and for identifying general principles that are insensitive to the details of a particular
inflationary model.

The relation between cosmological correlators and scattering amplitudes requires some
care. In a time-dependent cosmological background, the absence of the usual asymptotic
in- and out-states prevents the definition of a conventional S-matrix. Instead, one studies
two closely related boundary quantities: coefficients of the late-time wavefunction and
equal-time correlators computed in the in--in formalism. Wavefunction coefficients are the
closer analogues of scattering amplitudes. Their singularities and canonical forms can be
described using cosmological polytopes~\cite{Arkani-Hamed:2017fdk}, their residue at
vanishing total energy reproduces the corresponding flat-space
amplitude~\cite{Maldacena:2011nz,Raju:2012zr}, and they obey simple systems of
differential equations~\cite{Arkani-Hamed:2023bsv,Arkani-Hamed:2023kig,De:2023xue,Baumann:2024mvm,He:2024olr}.
Cosmological correlators, by contrast, are the late-time expectation values obtained from
the squared modulus of the wavefunctional. They depend on spatial momenta rather than
Lorentz-invariant Mandelstam variables, with momentum magnitudes and their sums playing
the role of energies. Although these energies are not subject to a conservation law, the
analytic continuation to vanishing total energy again exposes the corresponding
flat-space scattering data.

At first sight, correlators might be expected to have a more complicated analytic
structure than wavefunction coefficients, since they are assembled from sums of products
of the latter. Recent results reveal a different pattern. After the cancellations inherent in
the in--in combination are taken into account, correlators can have lower transcendental
weight and smaller symbol alphabets than the corresponding wavefunction
coefficients~\cite{Chowdhury:2023arc,Chowdhury:2025ohm,Chowdhury:2026upp,Chowdhury:2026dwm}.
This distinction is already striking at one loop. For the graphs relevant here, the
two-site one-loop wavefunction is polylogarithmic, whereas the three-site one-loop
wavefunction involves elliptic functions~\cite{Benincasa:2024ptf}. It is therefore natural
to ask how much of this apparent complexity survives in the corresponding correlator and
whether its simplification can be understood directly from an integral representation.

In this work, we investigate this question for conformally coupled scalars in $\mathrm{dS}_4$ with $\phi^4$ self-interactions. Their mode functions are plane waves up to a simple power
of conformal time, making it possible to reduce the time integrations to
flat-space-type energy integrals while retaining the characteristic analytic structure of a
cosmological correlator. Although this theory is not itself a realistic model of primordial
fluctuations, it provides a useful seed for more physical observables. At tree level,
correlators of the curvature perturbation can in suitable cases be related to conformally
coupled seed correlators by weight-shifting operators and soft
limits~\cite{Baumann:2019oyu}. Whether an equally systematic construction persists for
loop-level in--in correlators remains an open question. Nevertheless, compact loop-level
seed correlators provide a controlled setting in which to identify the singularity,
factorisation, and contour structures that may survive in more realistic theories.

The particular object studied in this paper is the one-loop three-site correlator, or
``triangle'', with two external legs attached to each quartic vertex. It is the first case in
which the loop integration is both nontrivial and finite. A closed-form evaluation was
recently obtained in ref.~\cite{Pimentel:2026kqc}. That work introduced a set of reduced
integrals, analysed their singularities using Landau and Euler-discriminant methods, and
expressed the result in terms of dilogarithms. The calculation established that the
correlator is polylogarithmic despite the elliptic structure of the corresponding
wavefunction coefficient. The resulting representation, however, obscures this simplicity:
it contains one hundred and twenty dilogarithms, whose arguments involve two square
roots for each permutation; it requires an auxiliary regulator that must be removed in a
prescribed order relative to the $i\varepsilon$ prescription; and individual terms contain
spurious singularities that cancel only after the full expression is assembled.

This paper makes two main contributions. The first is a compact analytic representation
of the triangle correlator, obtained using methods developed for scattering amplitudes. We
organise the answer into six algebraic leading-singularity prefactors multiplying pure
functions of uniform transcendental weight two, and show that the leading singularities
are separately conformally invariant. Computing and simplifying the symbol in the sense
of~\cite{Goncharov:2010jf}, and subsequently integrating it back to a function, gives an
expression containing forty-two dilogarithms in place of the original one hundred and
twenty. All dilogarithm arguments are real in the physical region, each pure function
involves only one quadratic extension, and no auxiliary regulator is required.

The second main contribution is an independent derivation of the symbol directly from
the dressed integral representation. Pinches of the auxiliary-energy contours determine
the physical first entries, while localising the discontinuities across the corresponding
cuts determines the entries following them. This reconstructs the complete symbol without
first performing the two dressing integrations. We then show that still less detailed
information is sufficient: once the leading singularities, the Landau loci, and the
candidate algebraic letters are known, integrable condition, the $S_3$ permutation symmetry, and
the conformal Ward identities fix the symbol uniquely. Since none of these ingredients is
specific to the triangle, the same strategy provides a promising route towards higher-point
one-loop correlators. As a further structural result, we show that, after division by the
product of the three vertex energies, the correlator is a Stieltjes function of each squared
energy separately and is jointly completely monotone in all six squared energies.

\begin{figure}
\centering
\begin{tikzpicture}[scale=0.62,line width=0.8pt,
  inarrow/.style={postaction={decorate},
    decoration={markings,mark=at position 0.55 with {\arrow{Stealth}}}},
  lbl/.style={font=\small}]
\coordinate (V1) at (0,2.9);
\coordinate (V2) at (-1.9,0);
\coordinate (V3) at (1.9,0);
\draw (V1)--(V2) node[lbl,midway,above left] {$y_3$};
\draw (V1)--(V3) node[lbl,midway,above right] {$y_2$};
\draw (V2)--(V3) node[lbl,midway,below=3pt] {$y_1$};
\draw[inarrow] ($(V1)+(115:1.55)$)--(V1);
\draw[inarrow] ($(V1)+(65:1.55)$)--(V1);
\draw[inarrow] ($(V2)+(175:1.55)$)--(V2);
\draw[inarrow] ($(V2)+(235:1.55)$)--(V2);
\draw[inarrow] ($(V3)+(5:1.55)$)--(V3);
\draw[inarrow] ($(V3)+(305:1.55)$)--(V3);
\node[lbl,above left=-3pt]  at ($(V1)+(115:1.55)$) {$\vec p_{11}$};
\node[lbl,above right=-3pt] at ($(V1)+(65:1.55)$)  {$\vec p_{12}$};
\node[lbl,left=2pt]         at ($(V2)+(175:1.55)$) {$\vec p_{21}$};
\node[lbl,below left=-3pt]  at ($(V2)+(235:1.55)$) {$\vec p_{22}$};
\node[lbl,right=2pt]        at ($(V3)+(5:1.55)$)   {$\vec p_{31}$};
\node[lbl,below right=-3pt] at ($(V3)+(305:1.55)$) {$\vec p_{32}$};
\fill (V1) circle (2.2pt);
\fill (V2) circle (2.2pt);
\fill (V3) circle (2.2pt);
\node[lbl] at ($(V1)+(0,2.55)$) {$x_1=|\vec p_{11}|+|\vec p_{12}|$};
\node[lbl,anchor=north east] at ($(V2)+(-1.55,-1.55)$) {$x_2=|\vec p_{21}|+|\vec p_{22}|$};
\node[lbl,anchor=north west] at ($(V3)+(1.55,-1.55)$) {$x_3=|\vec p_{31}|+|\vec p_{32}|$};
\end{tikzpicture}
\caption{The one-loop three-site $\phi^4$ correlator. 
The two external momenta $\vec p_{i1},\vec p_{i2}$ entering vertex $i$ are drawn as incoming arrows, and the vertex carries the total energy $x_i=|\vec p_{i1}|+|\vec p_{i2}|$; the energies $y_1,y_2,y_3$ are
the magnitudes of the momenta injected at the vertices, and label the edges opposite them.}
\label{fig:graph}
\end{figure}
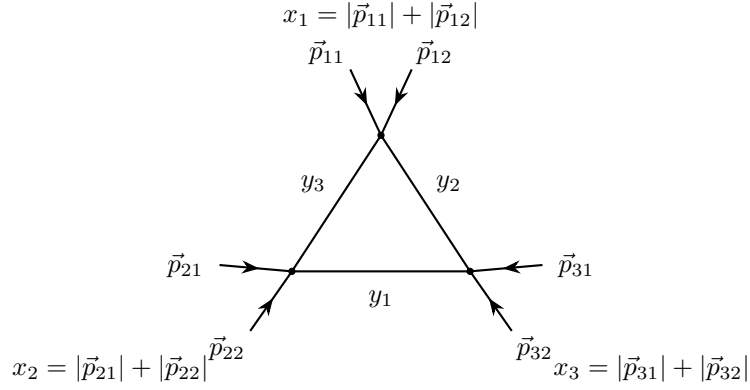

The paper is organised as follows. Section~\ref{sec:reps} reviews the kinematics and the
known representations of the triangle correlator. Section~\ref{sec:closed} presents the
compact analytic result, including the six leading singularities, the symbol alphabet, and
the function-level formula. Section~\ref{sec:checks} collects numerical and analytic
checks, including the conformal Ward identities and the relevant singular limits.
Section~\ref{sec:bootprog} derives the symbol directly from the dressed representation,
while section~\ref{sec:bootstrap} presents the complementary bootstrap based on Landau
singularities and consistency conditions. Appendix~\ref{app:positivity} establishes the Stieltjes and
complete-monotonicity properties and records the corresponding Feynman-parametric
representation.

\section{Review of known results}\label{sec:reps}

\subsection{Kinematics and conventions}\label{sec:notation}

We focus on conformally coupled scalars in dS$_4$ with $\phi^4$ interactions, as in~\cite{Chowdhury:2023arc,Pimentel:2026kqc}. The object of interest is the equal-time, one-loop correlator with the three-site
(triangle) topology, defined by the Schwinger--Keldysh (in--in) integral~\cite{Keldysh:1964ud},\footnote{\label{fn:norm}We strip off the couplings and the momentum-conserving delta function. We also strip off the universal external two-point-function factor $1/N$, where $N=\prod_{i=1}^{3}\prod_{\alpha=1}^{2}(2p_{i\alpha})$ and $p_{i\alpha}=|\vec p_{i\alpha}|$.
The normalisation is chosen such that
$\mathrm{Res}_{E_T=0}C$ is the off-shell flat-space triangle with unit coefficient~\eqref{eq:residue}.}
\begin{equation}\label{eq:inintriangle}
\begin{aligned}
C(x_i,y_i)
={}& \int\!\frac{\dd^3\ell}{(2\pi)^3}
\sum_{\sigma_1,\sigma_2,\sigma_3=\pm}
\prod_{i=1}^3(i\sigma_i)
\int_{-\infty}^{0}\!\prod_{i=1}^3\dd\eta_i\;
e^{i(x_1\eta_1+x_2\eta_2+x_3\eta_3)}
\\[-2pt]
&\quad\times
G_{\sigma_1\sigma_2}(\eta_1,\eta_2;|\vec\ell|)\,
G_{\sigma_2\sigma_3}(\eta_2,\eta_3;|\vec\ell+\vk_2|)\,
G_{\sigma_3\sigma_1}(\eta_3,\eta_1;|\vec\ell-\vk_1|)\,.
\end{aligned}
\end{equation}
The integral defines a six-point correlator in $\phi^4$ theory, with two external legs attached
to each quartic vertex (see Fig.~\ref{fig:graph}). Let $\vec p_{i1}$ and $\vec p_{i2}$ be the two external spatial momenta entering vertex $i$, for $i=1,2,3$. The kinematic dependence of the correlator is specified by
\begin{equation}\label{eq:external-kinematics}
x_i=|\vec p_{i1}|+|\vec p_{i2}|,\qquad
\vk_i=\vec p_{i1}+\vec p_{i2},\qquad
y_i=|\vk_i|,\qquad i=1,2,3.
\end{equation}
Here, $G_{\sigma\sigma'}$ are the Bunch--Davies Schwinger--Keldysh propagators. Because the fields are conformally coupled, the mode functions are
flat-space plane waves up to a power of the conformal time, so the propagators are
\begin{equation}
\begin{aligned}
&G_{\pm \pm}(\eta_1,\eta_2;Y):=\frac{1}{2Y}\left[ e^{\mp \text{i} \, Y (\eta_1-\eta_2)}\theta(\eta_1-\eta_2)+e^{\pm \text{i} \, Y (\eta_1-\eta_2)} \theta(\eta_2-\eta_1) \right],\\
& G_{\pm \mp}(\eta_1,\eta_2;Y):=\frac{1}{2Y} e^{\pm \text{i} \, Y (\eta_1-\eta_2)}\,.
\end{aligned}
\end{equation}
Performing the three nested integrations over the conformal times $\eta_i$ gives a rational function.
Subsequently performing the loop-momentum integration over $\ell$, the triangle contribution depends on the external kinematics through the six energies $x_i,y_i$ only. Our convention is related to that of ref.~\cite{Pimentel:2026kqc} by
\begin{equation}\label{eq:ident}
(x_1,x_2,x_3;y_1,y_2,y_3):=(X_1,X_2,X_3;k_1,k_2,k_3)\,.
\end{equation}
More precisely, ref.~\cite{Pimentel:2026kqc} computes equal-time in--in correlators of massless scalars in flat space; for conformally coupled scalars in dS$_4$, the two computations agree after the conformal rescaling of the fields by powers of the conformal time.
The $+++$ and $+-+$ terms in the sum reproduce Eqs.~(C.6) and~(C.7) of ref.~\cite{Pimentel:2026kqc}, up to the overall coupling constant.

Two important kinematic regions appear in this paper. The \emph{physical region} is the set of
kinematics that an actual correlator of on-shell external states can attain: the triangle
inequality gives $x_i\geq y_i$, while overall momentum conservation,
$\vk_1+\vk_2+\vk_3=\vec 0$, forces the three injected momenta to form a closed momentum
triangle. Explicitly,
\begin{equation}\label{eq:physregion}
x_i\ \ge\ y_i\ >\ 0\quad (i=1,2,3),\qquad y_i+y_j\ \ge\ y_k\,,
\end{equation}
where $i,j,k$ are pairwise distinct.

The \emph{Euclidean region} is the larger set $x_i\ge0$, $y_i\ge0$ for all six energies, still subject to
the second condition but with the constraints $x_i\ge y_i$ dropped. In its interior, $C$ is real and positive
(Appendix~\ref{app:positivity}). In particular, $C$ is real throughout the physical region, as observed in~\cite{Pimentel:2026kqc}; a general
proof that cosmological correlators are real at physical kinematics, from unitarity and scale invariance, is given in~\cite{Goodhew:2024eup}. For the unitarity constraints underlying such reality statements, see the cosmological optical theorem and the associated cutting rules~\cite{Goodhew:2020hob,Melville:2021lst,Goodhew:2021oqg}.

Outside the Euclidean region, the branch of each logarithm and dilogarithm must be specified, and we do so by the standard prescription of giving the vertex energies a small negative imaginary part,
$x_i\to x_i-i0$, inherited from the convergence of the conformal time integrals.

We introduce the following notation. We define the K\"all\'en function by
\begin{equation}
    \lambda(a,b,c)=a^2+b^2+c^2-2ab-2bc-2ca\,,
\end{equation}
which serves as the prototype for the radicands of all algebraic functions. In particular, we denote the Heron combination of the internal energies by
\begin{equation}\label{eq:p}
\begin{aligned}
p=-\lambda(y_1^2,y_2^2,y_3^2)
&=(y_1{+}y_2{+}y_3)(-y_1{+}y_2{+}y_3)(y_1{-}y_2{+}y_3)(y_1{+}y_2{-}y_3)\,.
\end{aligned}
\end{equation}
Finally, we also introduce the shorthand notation
\begin{equation}
 E_T=x_1+x_2+x_3,\quad
 u_1=-x_1+x_2+x_3,\quad
 u_2=x_1-x_2+x_3,\quad
 u_3=x_1+x_2-x_3.
\end{equation}

Another important object that appears throughout is the four-dimensional triangle with massless internal propagators,
\begin{equation}\label{eq:Tri}
\Tri(P_1^2,P_2^2,P_3^2)=\int\frac{{\rm d^4}L}{\pi^2}\frac{1}{L^2(L+P_1)^2(L+P_1+P_2)^2}\,,
\end{equation}
with all three legs off shell and $P_i^2>0$. We use the standard variables for the
off-shell massless triangle:
\begin{equation}\label{eq:flatzbar}
 z=
 \frac{P_1^2+P_3^2-P_2^2+\text{i}\sqrt{-\Delta_3}}{2P_3^2},\quad \bar{z}=
 \frac{P_1^2+P_3^2-P_2^2-\text{i}\sqrt{-\Delta_3}}{2P_3^2}
\end{equation}
where $\Delta_3=\lambda(P_1^2,P_2^2,P_3^2)$ and
$P_3={-}P_1{-}P_2$. For real momenta obeying momentum
conservation, one has $\Delta_3=-4\left[
P_1^2P_2^2-(P_1\cdot P_2)^2\right]\leq 0$.
The inequality is strict for non-collinear momenta, while equality corresponds to the degenerate boundary. In the
non-degenerate Euclidean region, the classic evaluation of the triangle~\cite{Usyukina:1992jd} can be written in
the manifestly real form
\begin{equation}\label{eq:TriUD}
\Tri(P_1^2,P_2^2,P_3^2)
=
\frac{1}{\sqrt{-\Delta_3}}
\biggl[
4\operatorname{Im}\text{Li}_2(z)
+4\log|z|\,\arg(1-z)
\biggr].
\end{equation}
The degenerate case $\Delta_3=0$ is obtained by continuity. In what follows, we also denote the function in square brackets by $\pT(P_1^2,P_2^2,P_3^2)$.

\subsection{The cut representation}\label{sec:maxcut}

The correlator~\eqref{eq:inintriangle} was first computed in~\cite{Pimentel:2026kqc}. Our main aim in
this work is to recast it in a form that is both much shorter and manifestly
organised by its singularities. In this subsection, we first review their result.

Let $\Om$ be the Gram determinant of the cut momenta $L_1=(-u,\vk_3)$, $L_2=(u-v,\vk_2)$,
$L_3=(v,\vk_1)$, so that $\Om=\lambda(L_1^2,L_2^2,L_3^2)$. Explicitly,
\begin{equation}\label{eq:Omega}
\Om(u,v)=y_1^4+y_2^4+y_3^4-2y_1^2y_2^2-2y_1^2y_3^2-2y_2^2y_3^2
-4uv\,(y_1^2-y_2^2+y_3^2)+4y_1^2u^2+4y_3^2v^2 ,
\end{equation}
and let $R=\{u\ge y_3,\ v\ge y_1,\ |u-v|\le y_2\}$. Defining
\begin{equation}\label{eq:Tab}
\mathcal T(a,b)=\int_R\frac{\dd u\,\dd v}{\sqrt{\Om(u,v)}\,(a+u)(b+v)} ,
\end{equation}
the correlator is the sum over the six permutations of $S_3$,
\begin{equation}\label{eq:Cmaxcut}
C=\sum_{\sigma\in S_3}\left[\frac{\mathcal T(x_3,x_2+x_3)}{E_T}
+\frac{\mathcal T(x_3,x_1)-\mathcal T(x_3,x_2+x_3)}{u_1}\right]_\sigma .
\end{equation}
This representation
is due to~\cite{Pimentel:2026kqc}: the region and the polynomial $\Om$ are their (2.36) and (2.38), and the
assembly~\eqref{eq:Cmaxcut} is their (4.48) and (4.49).\footnote{Eq. (2.38) in reference~\cite{Pimentel:2026kqc} contains a typo. The cross term should read
$-4uv(k_1^2-k_2^2+k_3^2)$ as in~\eqref{eq:Omega}, as required by the Gram determinant of the cut
momenta.}
This expression makes the observed weight-two structure natural: each $\mathcal{T}(a,b)$ is a double
integral of a weight-zero algebraic function, and the maximal cut in $(u,v)$ gives a unique maximal residue
\begin{equation}
    \frac{1}{\sqrt{\Omega(a,b)}}\radd{\,.}
\end{equation}
Together with the rational factors $1/E_T$ and $1/u_i$, this provides the starting point for our leading-singularity analysis in section~\ref{sec:sixLS}.

After performing the $S_3$ sum and grouping the result according to the square roots $\sqrt{\Omega}$, one obtains
\begin{equation}\label{eq:Tconn}
\mathcal T_1=\mathcal T(x_3,x_2{+}x_3)\big|_{(y_1,y_2,y_3)}
+\mathcal T(x_2,x_2{+}x_3)\big|_{(y_1,y_3,y_2)}\,,
\end{equation}
and cyclically for $\mathcal T_2,\mathcal T_3$, together with
\begin{equation}\label{eq:Tfold}
\mathcal T_4=\mathcal T(x_3,x_1)\big|_{(y_1,y_2,y_3)}\,,\qquad
\mathcal T_5=\mathcal T(x_2,x_1)\big|_{(y_1,y_3,y_2)}\,,\qquad
\mathcal T_6=\mathcal T(x_3,x_2)\big|_{(y_2,y_1,y_3)}\,,
\end{equation}
the subscript indicating the assignment of the internal energies in~\eqref{eq:Omega}. Each of
$\mathcal T_4,\mathcal T_5,\mathcal T_6$ appear twice in~\eqref{eq:Cmaxcut}, once with each of
the two partial-energy denominators.

The integrated result for each block $\mathcal{T}(a,b)$ contains 10 dilogarithms and 52 products of logarithms, and the complete expression as given contains 120 dilogarithms. It is written in terms of arguments involving two square roots per permutation, and
depends on an auxiliary regulator $\delta$ which is to be sent to zero at the end, jointly with the
$i\varepsilon$ of the energies and in a definite order.

\subsection{The dressed representation}\label{sec:dressed}
The cosmological dressing rules~\cite{Chowdhury:2023arc,Chowdhury:2025ohm,Chowdhury:2026upp} prescribe taking
the corresponding flat-space Feynman diagram, relaxing energy conservation, attaching one auxiliary
one-dimensional propagator to each vertex, joining the auxiliary propagators at a common point where energy is conserved, and integrating
over the auxiliary energies $p_i$. For the triangle, this leaves 
two integrations. The integral in~\cite{Chowdhury:2023arc} gives 
\begin{equation}\label{eq:Cdressed}
C=\frac{x_1x_2x_3}{\pi^2}\int_{\mathbb R^2}\!\dd p_1\,\dd p_2\;
\frac{\Tri\big(p_1^2+y_1^2,\;p_2^2+y_2^2,\;p_3^2+y_3^2\big)}
{\big(p_1^2+x_1^2\big)\big(p_2^2+x_2^2\big)\big(p_3^2+x_3^2\big)}\,,
\qquad -p_3=p_1+p_2 .
\end{equation}
The function $\Tri$ is the four-dimensional flat-space triangle~\eqref{eq:Tri}, evaluated with Euclidean four-vectors $P_i^E=(p_i,\vk_i)$. The momenta conservation condition is satisfied by the three momenta $P_i^E$, therefore
\begin{equation}\label{eq:lemmabound}
-\lambda\big(p_1^2+y_1^2,\;p_2^2+y_2^2,\;p_3^2+y_3^2\big)\;\ge\;0 \qquad\text{for all real }p_1,p_2 ,
\end{equation}
Especially, $-\lambda=p$ at $p_1=p_2=0$. The triangle in~\eqref{eq:Cdressed} never leaves the regime
$\lambda<0$, therefore the
integrand of~\eqref{eq:Cdressed} is real-analytic, with no cancellations anywhere.

Equation~\eqref{eq:Cdressed} is the Euclidean representation of the dressed
integral. For the analysis of poles and discontinuities later, it is also useful to
analytically continue the auxiliary contours back to real Lorentzian energies
$q_i$, equivalently $p_i=-iq_i$. The dressed representation then becomes \cite{Chowdhury:2026upp}
\begin{equation}\label{eq:CdressedLorentzian}
C=\frac{x_1x_2x_3}{\pi^2}
\int_{\mathbb R^2}\!\dd q_1\,\dd q_2\;
\frac{
\Tri\bigl(
q_1^2-y_1^2+i0,\,
q_2^2-y_2^2+i0,\,
q_3^2-y_3^2+i0
\bigr)
}{
\bigl(q_1^2-x_1^2+i0\bigr)
\bigl(q_2^2-x_2^2+i0\bigr)
\bigl(q_3^2-x_3^2+i0\bigr)
}\,,
\end{equation}
where $-q_3=q_1+q_2$. The $i0$ prescriptions specify the sides of the contour on which the
auxiliary poles and the branch points of the triangle lie, which we suppress in what follows. This continuation
changes only the auxiliary integration contours and gives the same correlator
$C$, not a change in the spacetime signature of the cosmological correlator. In the following, we use the Euclidean form~\eqref{eq:Cdressed} for numerical evaluation and
positivity, and the Lorentzian form~\eqref{eq:CdressedLorentzian} to study the
pole and pinch structure. Closely related reconstructions of cosmological correlators from single-cut discontinuities and dispersion relations, connecting cutting and dressing rules, have recently appeared in refs.~\cite{Das:2025qsh,Das:2026vfv}.

\section{From the symbol to a compact analytic formula}\label{sec:closed}
In this section, we present the main result of this work: a compact analytic formula for the one-loop triangle cosmological correlator at both the symbol and function levels. The result is obtained directly by simplifying the expression in~\cite{Pimentel:2026kqc} using symbol calculus.

Before turning to the explicit calculation, we briefly review symbols and symbol letters for iterated integrals. More precisely, we consider the class of Chen iterated integrals with ${\rm d}\log$ kernels~\cite{Chen:1977oja}. A {\it pure} iterated integral of weight $w$ may be written as
\begin{equation}
\mathcal{F}^{(w)}(a_i)
\equiv
\int_{0<t_1<\cdots<t_w<1}
{\rm d}\log a_1(t_1)\cdots {\rm d}\log a_w(t_w).
\end{equation}
More general quantities may contain rational or algebraic kinematic prefactors multiplying such pure functions, such as the triangle correlator discussed in this paper. In this work, we will only need logarithms $\log(x)$ and their products, and the dilogarithm
\begin{equation}
\operatorname{Li}_2(x):=-\int_0^x\frac{{\rm d}t}{t}\log(1-t).
\end{equation}

The total differential of a weight-$w$ function $\mathcal{F}^{(w)}$ can be written in the form
\begin{equation}
{\rm d}\mathcal{F}^{(w)}=\sum_i\mathcal{F}_i^{(w-1)}{\rm d}\log x_i,
\end{equation}
from which its \textit{symbol}~\cite{Goncharov:2010jf,Duhr:2011zq} is defined recursively by
\begin{equation}
\mathcal{S}\left(\mathcal{F}^{(w)}\right)=\sum_i\mathcal{S}\left(\mathcal{F}_i^{(w-1)}\right)\otimes x_i.
\end{equation}
For a pure weight-$w$ function, its symbol can in general be written as a length-$w$ tensor,
\begin{equation}\label{eq:generalsymbol}
\mathcal{S}\left(\mathcal{F}^{(w)}\right)=\sum_I c_I\,x_1^I\otimes\cdots\otimes x_w^I,
\end{equation}
where the coefficients $c_I$ are numerical constants, which we take to be rational, and the entries $x_j^I$ are functions of the external kinematics referred to as \textit{symbol letters}. The set of all letters appearing in the symbol is called the \textit{symbol alphabet}.

At the function level, one and the same quantity may admit many equivalent representations, related by functional identities among logarithms and dilogarithms. The symbol trivialises these identities and thus provides a representation-independent algebraic invariant. This makes it a natural intermediate step for simplification: one first brings the symbol to a compact form, and then reconstructs a correspondingly compact function-level representation.

\subsection{Strategy for simplification}\label{sec:strategy}
To simplify the result of~\cite{Pimentel:2026kqc}, we proceed in four steps. A similar strategy was employed in the simplification of the two-loop six-point MHV scattering amplitude in $\mathcal{N}=4$ sYM theory~\cite{Goncharov:2010jf}.

Our first task is to identify and fully determine a linearly independent basis for the algebraic prefactors multiplying the pure uniform-weight functions. In the context of Feynman integrals, such prefactors are naturally associated with the leading singularities of the integral~\cite{Cachazo:2008vp}. To see how this structure arises here, note that the integrated result for $\mathcal{T}(a,b)$ in~\eqref{eq:Tab} is a pure weight-two function multiplied by an overall algebraic prefactor. Each term in~\eqref{eq:Cmaxcut} involves two square roots, one of which cancels upon summing over permutations. Organising the permutations according to the surviving square root leaves six algebraically independent roots, each accompanied by a rational factor. 
We therefore analyse separately the six terms associated with these independent leading singularities. This decomposition is presented in section~\ref{sec:sixLS}.

The second step is to compute and simplify the symbol of the six pure functions.  Working at the symbol level allows us to expose the underlying analytic structure while avoiding functional identities that may obscure it in a direct representation in terms of polylogarithms. Since the functions appearing here have transcendental weight two, their symbols can be obtained using the replacements
\begin{equation}
\mathcal{S}\bigl(\operatorname{Li}_2(x)\bigr)=-(1-x)\otimes x,
\qquad
\mathcal{S}\bigl(\log x \log y\bigr)=x\otimes y+y\otimes x.
\end{equation}
The resulting simplified symbols are presented in section~\ref{sec:symbols}. In particular, the symbol does not retain information about the choice of analytic branches of the underlying functions. Since the small regulators $\delta$ and $\epsilon$ in the expressions of~\cite{Pimentel:2026kqc} only encode the corresponding branch prescriptions and do not alter the symbol letters, they can be safely set to zero at the symbol level.

The third step is to integrate the simplified symbols back to functions, {\it i.e.}, to find a simple function-level representation that reproduces the same symbol. In particular, for symbols of weight two, this procedure is systematic. For an arbitrary weight-two symbol, we first decompose it into symmetric and antisymmetric parts,
\begin{equation}
\sum_{i,j} c_{i,j} a_i \otimes a_j
=
\frac{1}{2}\sum_{i,j} c_{i,j}\left(a_i \otimes a_j + a_j \otimes a_i\right)
+
\frac{1}{2}\sum_{i,j} c_{i,j}\left(a_i \otimes a_j - a_j \otimes a_i\right).
\end{equation}
The symmetric part can be integrated straightforwardly using products of logarithms,
\begin{equation}
\mathcal{S}\left(\frac{1}{2}\sum_{i,j} c_{i,j}\log a_i \log a_j\right)
=
\frac{1}{2}\sum_{i,j} c_{i,j}\left(a_i \otimes a_j + a_j \otimes a_i\right),
\end{equation}
so the only nontrivial task is to identify a function-level representation of the antisymmetric part. At weight two, antisymmetric symbols can be generated by the Rogers dilogarithm,
\begin{equation}
\mathcal{S}\left(\operatorname{Li}_2(x) + \frac{1}{2}\log x \log(1-x)\right)
=
\frac{1}{2}\left[x \otimes (1-x) - (1-x) \otimes x\right].
\end{equation}
Therefore, one needs to search for suitable arguments $x$ such that both $x$ and $1-x$ factorise multiplicatively in terms of the symbol letters $\{a_i\}$. For the triangle correlator, each of the six terms can be integrated into a sum of differences of dilogarithm pairs whose arguments are exchanged under a sign flip of the square root appearing in the corresponding leading singularity, 
with the arguments given in Eqs.~\eqref{eq:zconn1}--\eqref{eq:wfold4}, supplemented by products of logarithms as in~\eqref{eq:Phifun}.

The fourth step is to fix the beyond-the-symbol ambiguities. The symbol is insensitive to lower-weight terms multiplied by transcendental constants, for example
\begin{equation}
\mathcal{S}\bigl({\rm i}\pi\log W\bigr)=0,
\qquad
\mathcal{S}\bigl(\pi^2\bigr)=0.
\end{equation}
Such ambiguities usually have to be fixed carefully, for instance by numerical matching. Fortunately, for the triangle correlator they can be determined without any numerical input. First, requiring the functions to be real throughout the physical region rules out possible imaginary terms of the form ${\rm i}\pi\log W$ and fixes the branches term by term in the decomposition. Second, an additive constant proportional to $\pi^2$ is ``Galois-even'' under a sign flip of the square root associated with each leading singularity, whereas the corresponding contribution to the term $R_a\Phi_a$ is odd. Such terms are therefore excluded by imposing the required Galois-even parity of the full result $C$. The resulting function-level expressions are presented in section~\ref{sec:compact} and are checked against independent numerical evaluations in section~\ref{sec:numchecks}.

In summary, the final results of this paper are given by~\eqref{eq:master} and~\eqref{eq:Phifun}, with the ingredients in~\eqref{eq:Fconn1}--\eqref{eq:wfold4}. They express the full correlator in terms of dilogarithms with real arguments, with no auxiliary regulator, no $i\epsilon$ prescription, and no spurious algebraic square roots. The expressions are valid as written throughout the physical region. On the boundary strata of~\eqref{eq:physregion}, where individual radicands $\Sigma_a$ may vanish, the formula is understood in the limiting sense.

\subsection{Main formula and leading singularities}\label{sec:sixLS}
The full one-loop triangle correlator is
\begin{tcolorbox}
\begin{equation}
 C(x_1,x_2,x_3,y_1,y_2,y_3)=\sum_{a=1}^{6}R_a\Phi_a.
 \label{eq:master}
\end{equation}
\end{tcolorbox}
\noindent In the following, we display the data for $(R_1,\Phi_1)$ and $(R_4,\Phi_4)$.  The remaining
pairs $(R_a,\Phi_a)$ are obtained by applying the same substitutions to
both members of the pair. 
We have
\begin{equation}
 (R_2,\Phi_2)=\sigma(R_1,\Phi_1),\qquad
 (R_3,\Phi_3)=\sigma^2(R_1,\Phi_1),
\end{equation}
where
\begin{equation}\label{eq:sigmamap}
 \sigma:(x_1,x_2,x_3;y_1,y_2,y_3)
 \mapsto(x_2,x_3,x_1;y_2,y_3,y_1).
\end{equation}
The pairs $(R_5,\Phi_5)$ and $(R_6,\Phi_6)$ are obtained from
$(R_4,\Phi_4)$ through
\begin{equation}
\begin{aligned}\label{eq:map6}
 (x_1,x_2,x_3;y_1,y_2,y_3)&\mapsto
 (x_1,x_3,x_2;y_1,y_3,y_2),&&a=5,\\
 (x_1,x_2,x_3;y_1,y_2,y_3)&\mapsto
 (x_2,x_1,x_3;y_2,y_1,y_3),&&a=6.
\end{aligned}
\end{equation}
The six leading singularities are
\begin{equation}
 R_1=\frac{1}{\sqrt{\Sigma_1}}
 \left(\frac1{E_T}-\frac1{u_1}\right),
 \qquad
 R_4=\frac{1}{\sqrt{\Sigma_4}}
 \left(\frac1{u_1}+\frac1{u_3}\right).
\end{equation}
The branch of $\sqrt{\Sigma_a}$ is fixed by continuation from positive
Euclidean kinematics.  Under the Galois involution
$\sqrt{\Sigma_a}\mapsto-\sqrt{\Sigma_a}$, both $R_a$ and $\Phi_a$ are
odd, so their product is invariant.

The radicands are
\begin{align}
\Sigma_1={}&4x_2^2y_3^2{+}4x_3^2y_2^2
{-}4x_2x_3(y_1^2{-}y_2^2{-}y_3^2){-}p,\\
\Sigma_4={}&4x_1^2y_3^2{+}4x_3^2y_1^2
{-}4x_1x_3(y_1^2{-}y_2^2{+}y_3^2){-}p.
\end{align}
Here, $p$ is the combination of the internal energies defined
in~\eqref{eq:p}. The positivity and the zero loci of the radicands
can be seen directly by rewriting the two independent cases as
\begin{align}
\frac{\Sigma_1}{4y_2^2y_3^2}
={}&
\left(
\frac{x_2}{y_2}
-
\frac{x_3}{y_3}
\frac{y_1^2-y_2^2-y_3^2}{2y_2y_3}
\right)^2+
\left(
\frac{x_3^2}{y_3^2}-1
\right)
\left[
1-
\left(
\frac{y_1^2-y_2^2-y_3^2}{2y_2y_3}
\right)^2
\right],
\\[1ex]
\frac{\Sigma_4}{4y_1^2y_3^2}
={}&
\left(
\frac{x_1}{y_1}
-
\frac{x_3}{y_3}
\frac{y_1^2-y_2^2+y_3^2}{2y_1y_3}
\right)^2+
\left(
\frac{x_3^2}{y_3^2}-1
\right)
\left[
1-
\left(
\frac{y_1^2-y_2^2+y_3^2}{2y_1y_3}
\right)^2
\right].
\end{align}
All terms on the right-hand sides are non-negative in the physical
region. Moreover, in its interior, the strict inequalities
$x_i>y_i$ and the strict triangle inequalities imply
$\Sigma_1>0$ and $\Sigma_4>0$. The remaining radicands obey the same
property by symmetry.

On the closure of the physical region, the two representative zero
loci are
\begin{align*}
\Sigma_1=0
&\quad\Longleftrightarrow\quad
y_1=y_2+y_3,
\qquad
\frac{x_2}{y_2}=\frac{x_3}{y_3},
\\
\Sigma_4=0
&\quad\Longleftrightarrow\quad
y_2=\lvert y_1-y_3\rvert,
\qquad
\frac{x_1}{y_1}=\frac{x_3}{y_3},
\end{align*}
together with their symmetry images. Thus, the radicands can vanish
only on degenerate collinear boundaries of the physical region. At a generic point of such a boundary, the function
$\Phi_a$ is locally regular in $\sqrt{\Sigma_a}$. Since it is odd
under the Galois involution
$\sqrt{\Sigma_a}\mapsto-\sqrt{\Sigma_a}$, its local expansion takes
the form
\begin{equation*}
    \Phi_a
    =
    \sqrt{\Sigma_a}\,\phi_a^{(0)}
    +\mathcal{O}\!\left(\Sigma_a^{3/2}\right),
\end{equation*}
where $\phi_a^{(0)}$ is finite. This cancels the explicit
$1/\sqrt{\Sigma_a}$ in $R_a$, so that $R_a\Phi_a$ has a finite limit.
Therefore, the vanishing of $\Sigma_a$ does not by itself produce a
physical singularity.

\subsection{The symbols and the alphabet}
\label{sec:symbols}\label{sec:symbol}

We first present the results for $\Phi_i$ at the symbol level:
\begin{tcolorbox}
\begin{equation}\label{eq:symres}
\cS(\Phi_a)=
\begin{cases}
\displaystyle\sum_{i=1}^{3}F^{(a)}_i\ox\chi^{(a)}_i\,, & a=1,2,3\,,\\[2.2ex]
\displaystyle\sum_{i=1}^{4}G^{(a)}_i\ox\upsilon^{(a)}_i\,, & a=4,5,6\,.
\end{cases}\;
\end{equation}
\end{tcolorbox}
\noindent The first-entry ratios are
\begin{equation}\label{eq:Fconn1}
\big\{F^{(1)}_i\big\}_{i=1,2,3}=\left\{
\frac{(x_2+y_1+y_3)(x_3+y_1+y_2)}{(x_2+x_3+y_1)(y_1+y_2+y_3)}\,,\
\frac{x_3+y_1+y_2}{x_3+y_3}\,,\
\frac{x_2+y_1+y_3}{x_2+y_2}\right\},
\end{equation}
\begin{equation}\label{eq:Gfold4}
\big\{G^{(4)}_i\big\}_{i=1,\dots,4}=\left\{
\frac{y_1+y_2+y_3}{x_3+y_1+y_2}\,,\
\frac{x_1+y_1}{x_3+y_1+y_2}\,,\
\frac{x_1+y_2+y_3}{x_3+y_3}\,,\
\frac{x_3+y_1+y_2}{x_1+y_2+y_3}\right\}.
\end{equation}
The corresponding last-entry ratios are
\begin{equation}\label{eq:chi1}
\begin{aligned}
\big\{\chi^{(1)}_i\big\}_{i=1,2,3}=\bigg\{&
\frac{2x_2x_3-y_1^2+y_2^2+y_3^2+\sqrt{\Sigma_1}}{2x_2x_3-y_1^2+y_2^2+y_3^2-\sqrt{\Sigma_1}}\,,\\
&\frac{(y_1+y_2)^2-y_3^2+2x_2y_1+2x_2y_2+2x_3y_2+\sqrt{\Sigma_1}}{(y_1+y_2)^2-y_3^2+2x_2y_1+2x_2y_2+2x_3y_2-\sqrt{\Sigma_1}}\,,\\
&\frac{(y_1+y_3)^2-y_2^2+2x_2y_3+2x_3y_1+2x_3y_3+\sqrt{\Sigma_1}}{(y_1+y_3)^2-y_2^2+2x_2y_3+2x_3y_1+2x_3y_3-\sqrt{\Sigma_1}}\bigg\}\,,
\end{aligned}
\end{equation}
and
\begin{equation}\label{eq:ups4}
\begin{aligned}
\big\{\upsilon^{(4)}_i\big\}_{i=1,\dots,4}=\bigg\{&
\frac{-2x_1x_3+y_1^2-y_2^2+y_3^2+\sqrt{\Sigma_4}}{-2x_1x_3+y_1^2-y_2^2+y_3^2-\sqrt{\Sigma_4}}\,,\\
&\frac{y_1^2-(y_2+y_3)^2+2x_1y_3-2x_3y_2-2x_3y_3+\sqrt{\Sigma_4}}{y_1^2-(y_2+y_3)^2+2x_1y_3-2x_3y_2-2x_3y_3-\sqrt{\Sigma_4}}\,,\\
&\frac{(y_1+y_2)^2-y_3^2-2x_3y_1+2x_1y_1+2x_1y_2+\sqrt{\Sigma_4}}{(y_1+y_2)^2-y_3^2-2x_3y_1+2x_1y_1+2x_1y_2-\sqrt{\Sigma_4}}\,,\\
&\frac{(y_1-y_3)^2-y_2^2-2x_3y_1+2x_1y_3+\sqrt{\Sigma_4}}{(y_1-y_3)^2-y_2^2-2x_3y_1+2x_1y_3-\sqrt{\Sigma_4}}\bigg\}\,.
\end{aligned}
\end{equation}
The remaining blocks follow from the substitutions~\eqref{eq:sigmamap}--\eqref{eq:map6}.

\subsection{The final function-level result}
\label{sec:compact}

Finally, we present the function-level result obtained through the integration procedure described in section~\ref{sec:strategy}. To present our result compactly, we introduce a single real function of one variable:
\begin{equation}\label{eq:Ldef}
L(z)=\begin{cases}
\mathrm{Li}_2(z)+\tfrac12\log z\,\log(1-z)\,, & 0<z\le1\,,\\[0.4ex]
-L\big(\tfrac{z}{z-1}\big)\,, & z\le0\,,\\[0.4ex]
-L(1/z)-\tfrac{\pi^2}{6}\,, & z>1\,,
\end{cases}
\end{equation}
which gives a real-valued, piecewise prescription for the Rogers dilogarithm on the entire real axis, starting from $(0,1]$ where it is real. For the inequivalent real continuations of the Rogers dilogarithm beyond this interval, see ref.~\cite{Zagier:2007knq}. The role of this particular prescription is explained in section~\ref{sec:continuity}. Using this shorthand, the compact function-level result is
\begin{tcolorbox}[left=1mm,right=1mm]
\begin{equation}\label{eq:Phifun}
\Phi_a=
\begin{cases}
\displaystyle-\sum_{i=1}^{3}\Big[L\big(z^{(a)}_i\big)-L\big(\bar z^{(a)}_i\big)\Big]
+\frac14\sum_{i=1}^{3}\log F^{(a)}_i\,\log\Big[\big(\chi^{(a)}_i\big)^2\Big]\,, & a=1,2,3\,,\\[2.6ex]
\displaystyle-\sum_{i=1}^{4}\Big[L\big(w^{(a)}_i\big)-L\big(\bar w^{(a)}_i\big)\Big]
+\frac14\sum_{i=1}^{4}\log G^{(a)}_i\,\log\Big[\big(\upsilon^{(a)}_i\big)^2\Big]\,, & a=4,5,6\,,
\end{cases}
\end{equation}
\end{tcolorbox}
\noindent with the first- and last-entry ratios of section~\ref{sec:symbols}. The dilogarithm
arguments are
\begin{align}
z^{(1)}_1&=\frac{(y_1-y_2+y_3)\big(2x_2x_3-y_1^2+y_2^2+y_3^2+\sqrt{\Sigma_1}\big)}
{(x_2+y_2)\big[(y_1+y_3)^2-y_2^2+2x_2y_3+2x_3y_1+2x_3y_3-\sqrt{\Sigma_1}\big]}\,,\label{eq:zconn1}\\
z^{(1)}_2&=-\frac{2x_2x_3-y_1^2+y_2^2+y_3^2+\sqrt{\Sigma_1}}
{(y_1+y_2)^2-y_3^2+2x_2y_1+2x_2y_2+2x_3y_2-\sqrt{\Sigma_1}}\,,\label{eq:zconn2}\\
z^{(1)}_3&=-\frac{(y_1+y_2)^2-y_3^2+2x_2y_1+2x_2y_2+2x_3y_2+\sqrt{\Sigma_1}}
{(y_1+y_3)^2-y_2^2+2x_2y_3+2x_3y_1+2x_3y_3-\sqrt{\Sigma_1}}\,,\label{eq:zconn3}
\end{align}
and
\begin{align}
w^{(4)}_1&=\frac{\sqrt{\Sigma_4}-2x_1y_1+2x_3y_1-y_1^2-2x_1y_2-2y_1y_2-y_2^2+y_3^2}
{\sqrt{\Sigma_4}+2x_1^2-2x_1x_3-y_1^2-y_2^2+y_3^2}\,,\label{eq:wfold1}\\
w^{(4)}_2&=\frac{\sqrt{\Sigma_4}+2x_1x_3-y_1^2+y_2^2-y_3^2}
{\sqrt{\Sigma_4}-y_1^2-2x_3y_2+y_2^2+2x_1y_3-2x_3y_3+2y_2y_3+y_3^2}\,,\label{eq:wfold2}\\
w^{(4)}_3&=\frac{\sqrt{\Sigma_4}+2x_3y_1-y_1^2+y_2^2-2x_1y_3+2y_1y_3-y_3^2}
{\sqrt{\Sigma_4}+2x_1x_3-y_1^2+y_2^2-y_3^2}\,,\label{eq:wfold3}\\
w^{(4)}_4&=\frac{\sqrt{\Sigma_4}+2x_1x_3-2x_3^2-y_1^2+y_2^2+y_3^2}
{\sqrt{\Sigma_4}-2x_1y_1+2x_3y_1-y_1^2-2x_1y_2-2y_1y_2-y_2^2+y_3^2}\,.\label{eq:wfold4}
\end{align}
Each argument is chosen such that both the argument itself and one minus the argument factorise multiplicatively in terms of the factors appearing in the symbol letters~\eqref{eq:Fconn1}--\eqref{eq:ups4}, as required by the integration procedure of section~\ref{sec:strategy}. The remaining terms follow from the substitutions~\eqref{eq:sigmamap}--\eqref{eq:map6}.
For each dilogarithm pair, we define Galois conjugates by
\begin{equation}\label{eq:galois}
\bar z^{(a)}_i:=z^{(a)}_i\big|_{\sqrt{\Sigma_a}\to-\sqrt{\Sigma_a}}\,,\qquad
\bar w^{(a)}_i:=w^{(a)}_i\big|_{\sqrt{\Sigma_a}\to-\sqrt{\Sigma_a}}\,,
\end{equation}
so every pair is odd, as $\Phi_a$ must be.

Equations~\eqref{eq:master},~\eqref{eq:Fconn1}--\eqref{eq:ups4} and~\eqref{eq:Phifun}--\eqref{eq:wfold4} are the main result of this paper. Owing to the piecewise definition of $L(z)$ and the squared argument of the second logarithm in $\Phi_a$, all arguments are real, and each dilogarithm pair and logarithm product is individually real throughout the physical region. Consequently, the final expression requires neither of the regulators $\delta$ and $\epsilon$ appearing in the original representation of \cite{Pimentel:2026kqc}.

\section{Checks and analytic properties}\label{sec:checks}
The compact formula obtained in the previous section can be tested in three complementary ways. We first compare it numerically with the
maximal-cut and dressed representations over the physical region. We then study the conformal Ward identities, which provide analytic constraints on the leading singularities and on the pairing of symbol
entries. Finally, we examine its singular limits: the total-energy residue reproduces the flat-space triangle, the partial-energy limit exhibits the expected factorisation, and the apparent poles of the rational prefactors cancel. We also discuss the continuity properties of the building blocks $\Phi_a$ (section~\ref{sec:continuity}). Besides providing checks of the result, these properties constitute part of the
input for the bootstrap developed in section~\ref{sec:bootstrap}.

\subsection{Numerical checks}
\label{sec:numchecks}

We validate the compact formula through three complementary comparisons.
The reference values are obtained by evaluating the closed
form~\eqref{eq:master}, with the ingredients in
\eqref{eq:Fconn1}--\eqref{eq:ups4} and the blocks in
\eqref{eq:Phifun}--\eqref{eq:wfold4}, directly from the printed formulae
at $40$-digit precision.

The first comparison uses the analytic
representation $C_{\mathrm{PW}}$ from the ancillary file of \cite{Pimentel:2026kqc}. Table~\ref{tab:random} lists the
value of the compact formula at six physical reference points used for
this pointwise comparison. We evaluate the physical real part of the
result with the finite regulators
$\delta=10^{-9}$ and $\epsilon=10^{-11}$. It reproduces the reference
values with relative deviations ranging from $9\times10^{-12}$ to
$4\times10^{-10}$. These deviations decrease as the regulators are
reduced, indicating that they arise from the finite-regulator
prescription rather than from a difference between the two analytic
results.

\begin{table}
  \centering
  \begin{tabular}{lcc}
    \toprule
    $x_1,x_2,x_3;y_1,y_2,y_3$
    & $C$
    & $\left|\operatorname{Re}C_{\mathrm{PW}}/C-1\right|$ \\
    \midrule
    $3.55,\,2.76,\,2.50;\,1.15,\,0.93,\,1.00$
      & $0.074133734894611$
      & $1.29\times10^{-11}$ \\
    $3.10,\,2.20,\,4.00;\,0.70,\,0.90,\,1.10$
      & $0.023246429662069$
      & $9.42\times10^{-11}$ \\
    $4.20,\,3.30,\,2.80;\,1.40,\,1.10,\,0.80$
      & $0.150915767878964$
      & $4.88\times10^{-11}$ \\
    $2.90,\,3.70,\,3.40;\,0.85,\,1.25,\,1.05$
      & $0.076346805699096$
      & $2.13\times10^{-11}$ \\
    $5.10,\,2.40,\,3.60;\,1.30,\,0.75,\,1.20$
      & $0.046798497540527$
      & $3.62\times10^{-10}$ \\
    $3.80,\,4.10,\,5.00;\,1.05,\,1.35,\,0.95$
      & $0.022847266473144$
      & $9.34\times10^{-12}$ \\
    \bottomrule
  \end{tabular}
  \caption{Reference values of the compact formula \eqref{eq:master} at six physical
  points, used for comparison with the analytic
  result $C_{PW}$ in~\cite{Pimentel:2026kqc}. The coordinates are exact to
  the two decimal places shown. The final column gives the relative
  deviation of the physical real part of the previous representation
  from the compact formula, evaluated with $\delta=10^{-9}$ and
  $\epsilon=10^{-11}$.}
  \label{tab:random}
\end{table}

The second and third comparisons use, respectively, the cut
representation~\eqref{eq:Cmaxcut} and the dressed
representation~\eqref{eq:Cdressed}. For the cut representation, the
two-fold integrals of the type~\eqref{eq:Tab} are mapped to unit squares
and evaluated numerically after applying the required $S_3$
relabellings. For the dressed representation, the flat-space triangle is
integrated against the auxiliary propagators using iterated
double-exponential quadrature \cite{TakahasiMori:1973}, followed by a fourth-order Richardson
extrapolation \cite{Sidi:2003}. These two calculations involve different integration
domains and substantially different analytic inputs, and therefore
provide largely independent checks of the compact result.

\begin{figure}
  \centering
  \includegraphics[width=0.75\textwidth]{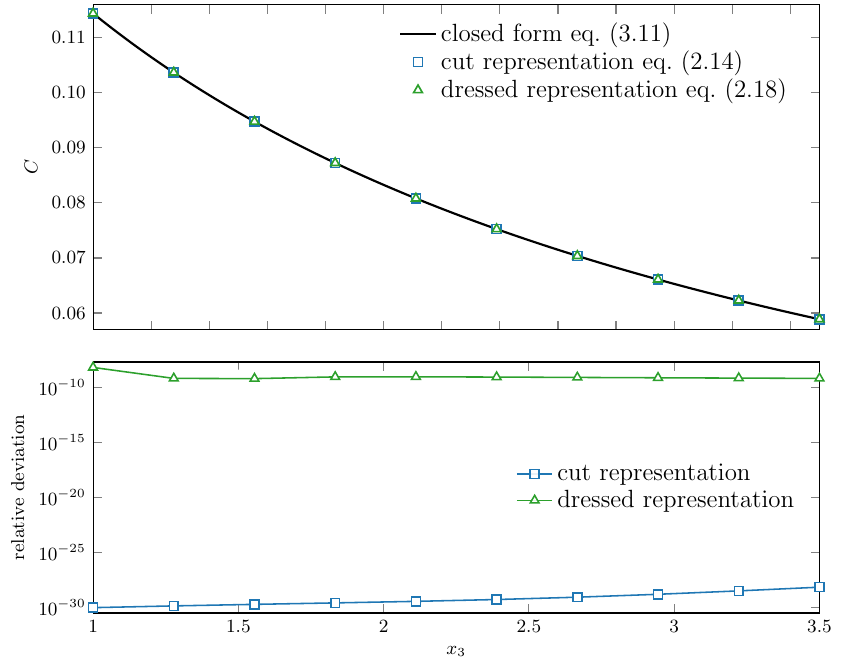}
  \caption{Comparison of the three evaluations of $C$.}
  \label{fig:four}
\end{figure}

Figure~\ref{fig:four} compares the compact formula with the cut and
dressed representations along
$x_3\in[1,7/2]$, with $x_1=3.549$, $x_2=2.764$, and $(y_1,y_2,y_3)=(1.15,0.93,1.00)$. The entire scan lies in the physical region, including the boundary point at $x_3=1$. The cut and dressed representations are both evaluated at
\begin{equation}
  x_3\in\left\{1,\frac{23}{18},\frac{14}{9},\frac{11}6,\frac{19}9,\frac{43}{18},\frac{8}3,\frac{53}{18},\frac{29}9,\frac{7}2\right\}.
\end{equation}
In the upper panel, the three evaluations are indistinguishable; the lower panel shows the relative deviations of the cut and dressed
representations from the closed form. At the ten sampled points, the cut representation
agrees with the compact formula to a relative accuracy of $7\times10^{-29}$ or better, while the dressed representation agrees to $7\times10^{-9}$ or better, with the precision limited by the corresponding numerical quadratures.

\subsection{Conformal Ward identities}
\label{sec:cwi}

The reduced correlator must satisfy the conformal Ward identities of the
late-time boundary theory~\cite{Arkani-Hamed:2018kmz}. In the following, we assume generic momenta, so as to avoid potential contact terms (see e.g. \cite{Chicherin:2017bxc,Chicherin:2022gky}).

Written in the six variables $x_i$ and $y_i$, the
dilatation Ward identity takes the form
\begin{equation}\label{eq:euler}
\left[
\sum_{i=1}^{3}
\left(x_i\,\partial_{x_i}+y_i\,\partial_{y_i}\right)+3
\right]C=0 .
\end{equation}
Thus $C$ is homogeneous of degree $-3$, as is manifest in the closed
formula.

The special conformal Ward identities require slightly more care because the
universal external two-point-function factors have been stripped from
$C$ (see footnote~\ref{fn:norm}). For one external leg, the scalar momentum-space generator obeys
\begin{equation}
\vec{\mathcal K}_1\!\left[\frac{1}{N}\,f\right]
=
\frac{1}{N}\,\vec{\mathcal K}_2 f,
\qquad
\vec{\mathcal K}_\Delta
:=
\vec p\,\bigl(\vec{\partial}_p\cdot\vec{\partial}_p\bigr)
-2\bigl(\vec p\cdot\vec{\partial}_p\bigr)\vec{\partial}_p
+2(\Delta-3)\vec{\partial}_p .
\end{equation}
Consequently, the Ward identity for the full six-point correlator is
equivalent to the $\Delta=2$ identity acting on the reduced correlator.
Summing the generators over the two external legs attached to vertex $i$
and applying the chain rule gives
\begin{equation}
\sum_{\alpha=1}^{2}\vec{\mathcal K}_{2,i\alpha}C
=
-\vk_i\,\mathcal D_i C \,,
\end{equation}
where
\begin{equation}
\mathcal D_i
\equiv
\partial_{x_i}^2+\partial_{y_i}^2
+\frac{2x_i}{y_i}\,\partial_{x_i}\partial_{y_i}
+\frac{2}{y_i}\,\partial_{y_i}.
\end{equation}
Using $\vec k_1+\vec k_2+\vec k_3=0$, the vector Ward identity is therefore
equivalent to
\begin{equation}\label{eq:cwipair}
\mathcal D_1 C=\mathcal D_2 C=\mathcal D_3 C,
\end{equation}
Substituting~\eqref{eq:master} into these equations, we verified the
Ward identities algebraically. 

A useful additional fact is that every
leading singularity satisfies them separately,
\begin{equation}\label{eq:LSannihilate}
\bigl(\mathcal D_i-\mathcal D_j\bigr)R_a=0,
\qquad
a=1,\ldots,6 .
\end{equation}
This is expected from cut-based arguments; see ref.~\cite{Henn:2021aco}. Analogous statements hold for scattering amplitudes, where conformally invariant leading singularities appear, for example, in two-loop six-gluon QCD amplitudes~\cite{Carrolo:2025agz}.
The remaining conditions therefore constrain the transcendental blocks
and their relative coefficients. At weight one, they become linear
relations among the allowed pairings of rational first entries and
Galois-odd last entries.

\subsection{Singular limits and factorisation}
\label{sec:residue}

The compact result also makes it straightforward to study the physically
important singular limits of the correlator. We first recover the flat-space triangle
from the total-energy pole, then verify the expected partial-energy
factorisation, and finally show that the additional poles visible in the
individual rational prefactors cancel in the full answer.

Only the three terms $a=1,2,3$ have a pole at
$E_T=x_1+x_2+x_3=0$; their residues are $1/\sqrt{\Sigma_a}$, whereas the
three terms  $a=4,5,6$ are regular. On this locus, we introduce the off-shell
invariants $P_i^2$ and the corresponding K\"all\'en discriminant. The three
quadratic extensions of the terms $a=1,2,3$ then degenerate to the same
extension:
\begin{equation}\label{eq:flatroot}
 P_i^2=x_i^2-y_i^2,\qquad
 \Delta_3=\lambda(P_1^2,P_2^2,P_3^2),\qquad
 \left.\Sigma_1\right|_{E_T=0}
 =
 \left.\Sigma_2\right|_{E_T=0}
 =
 \left.\Sigma_3\right|_{E_T=0}
 =
 \Delta_3 .
\end{equation}

After choosing a common branch of $\sqrt{\Delta_3}$, the three
symbols can therefore be combined. Before imposing $E_T=0$, the three terms contain nine tensor
terms. Relations among their rational first entries and algebraic second
entries reduce these. The resulting residue is
\begin{equation}\label{eq:residue}
\operatorname*{Res}_{E_T=0}
\left[\sum_{a=1}^{6}R_a\cS(\Phi_a)\right]
=
\frac{1}{\sqrt{\Delta_3}}
\left[
\frac{P_1^2}{P_3^2}\ox\frac{1-z}{1-\bar z}
+
\frac{P_2^2}{P_3^2}\ox\frac{\bar z}{z}
\right].
\end{equation}
This is precisely the symbol of the
four-dimensional massless triangle with three off-shell legs. The
$1/\sqrt{\Delta_3}$ factor is its leading singularity. Thus the
total-energy residue reproduces the expected flat-space symbol.

The answer also has the expected factorisation at
$x_2+x_3+y_1=0$. As discussed in ref.~\cite{Pimentel:2026kqc}, its leading
non-analytic behaviour is
\begin{equation}\label{eq:partialenergyfactorisation}
\begin{aligned}
\left.C\right|_{x_2+x_3+y_1\to0}
={}&
\left(
\frac{1}{x_1+y_1}+\frac{1}{x_1-y_1}
\right)
\log(x_2+x_3+y_1)
\\[+2pt]
&\times
\frac{\log(-P_2^2)-\log(-P_3^2)}
{P_3^2-P_2^2}
+O(1),
\end{aligned}
\end{equation}
where $O(1)$ denotes terms that remain finite in the limit. This
factorisation follows directly from our symbol: taking the discontinuity
in the first entry $x_2+x_3+y_1$ and then restricting to its vanishing
locus leaves the weight-one symbol of
$\log(-P_2^2)-\log(-P_3^2)$, with exactly the rational coefficient shown
in~\eqref{eq:partialenergyfactorisation}. 
Thus the discontinuity factorises into a shifted tree-level factor depending
on $x_1$ and $y_1$ and the flat-space-triangle discontinuity with one null
external invariant.

Finally, although some individual terms in the decomposition
\eqref{eq:master} have poles at $u_i=0$, these poles cancel in the full
sum, in agreement with the expected analytic properties of the
Bunch--Davies correlator~\cite{Arkani-Hamed:2018kmz}. At the symbol level,
this cancellation reads
\begin{equation}
\underset{u_i=0}{\operatorname{Res}}
\left[
\sum_{a=1}^{6}R_a\,\mathcal{S}(\Phi_a)
\right]
=0,
\qquad i=1,2,3.
\end{equation}
The cancellation involves different leading-singularity terms and
confirms that the $u_i$ poles introduced by the decomposition
\eqref{eq:master} are spurious.

\subsection{Continuity of the function}
\label{sec:continuity}

There is one important subtlety in the definition of $L(z)$:
for an individual term, crossing the discontinuity at $z=1$ shifts that
term by $\pm\pi^2/2$, since, according to the definition \eqref{eq:Ldef},
\begin{equation}
 \lim_{z\to1^-}L(z)-\lim_{z\to1^+}L(z)=\frac{\pi^2}{2}.
\end{equation}
At first sight, one may worry that our expressions are not continuous at the loci
$z_i^{(a)}=1$ and $w_i^{(a)}=1$. This apparent issue, however, does not
arise: each $\Phi_i$ is continuous inside the physical region.

To see analytically why the jumps of the individual $L(z)$ terms do not
affect the continuity of the $\Phi_i$, we take $\Phi_1$ and $\Phi_4$ as
examples. For the loci $z_1^{(1)}=1$ or $\bar z_1^{(1)}=1$, it suffices
to compute the product
\begin{equation}
\begin{aligned}
(z_1^{(1)}{-}1)(\bar z_{1}^{(1)}{-}1)
=\frac{
 (x_2+y_1+y_3)^2(y_1+y_2-y_3)
}{
 (x_2+y_2)(x_2+x_3+y_1)(y_1+y_2+y_3)
}.
\end{aligned}
\end{equation}
Every sum appearing on the right-hand side is positive in the physical
region, while the strict triangle inequality gives
$y_1+y_2-y_3>0$. Hence neither $z_1^{(1)}$ nor its Galois conjugate can
reach $1$ in the interior. The same reasoning applies to the other
$z_i^{(1)}$, as well as to $w_2^{(4)}$ and $w_3^{(4)}$.

The only case requiring additional care is the pair $w_1^{(4)}$ and
$w_4^{(4)}$, for which a similar computation gives
\begin{align}
\bigl(w^{(4)}_1-1\bigr)
\bigl(\bar w^{(4)}_1-1\bigr)
&=
\frac{
 (x_1+y_1)(x_1-x_3+y_2)
}{
 (x_1-y_1)(x_1-x_3-y_2)
},
\\[1ex]
\bigl(w^{(4)}_4-1\bigr)
\bigl(\bar w^{(4)}_4-1\bigr)
&=
\frac{
 (x_1-x_3+y_2)(x_3+y_1+y_2)^2
}{
 (x_1+y_1)(y_1+y_2-y_3)(y_1+y_2+y_3)
}.
\end{align}
Both products therefore vanish on the same locus $x_3=x_1+y_2$.
However, on this locus, the radicand becomes a perfect square,
\begin{equation}
\left.\Sigma_4\right|_{x_3=x_1+y_2}
=
\left(y_1^2+2x_1y_2+y_2^2-y_3^2\right)^2.
\end{equation}
Choosing the branch of $\sqrt{\Sigma_4}$ by continuation from the physical region
then gives
\begin{equation}
\bar w^{(4)}_1=\bar w^{(4)}_4=1.
\end{equation}
Thus, the two arguments $\bar w_1^{(4)}$ and $\bar w_4^{(4)}$ reach the
discontinuity point of $L(z)$ simultaneously, while $w_1^{(4)}$ and
$w_4^{(4)}$ do not reach $1$ on this sheet. Finally, it can be checked
that the two arguments obey the relation
\begin{equation}
\bar w^{(4)}_1-1
=
-\frac{x_1+y_1}{x_3+y_1+y_2}\,
\bar w^{(4)}_1
\bigl(\bar w^{(4)}_4-1\bigr).
\end{equation}
Since both the ratio $(x_1+y_1)/(x_3+y_1+y_2)$ and $\bar w_1^{(4)}$ are positive in
a neighbourhood of this locus, $\bar w^{(4)}_1-1$ and
$\bar w^{(4)}_4-1$ have opposite signs, so the arguments
$\bar w^{(4)}_1$ and $\bar w^{(4)}_4$ cross $1$ in opposite directions. Their two $\pi^2/2$ jumps therefore cancel, leaving
$\Phi_4$ continuous at this locus.

\section{From the dressed representation to the symbol}\label{sec:bootprog}
In this section, we show that the symbol in~\eqref{eq:symres} can be
derived independently from the dressed representation~\eqref{eq:Cdressed}.
First, we decompose the dressed integrand into the six pure sectors
associated with the leading singularities. Second, we determine the
first entries by analysing the physical contour pinches. Third, we
compute the discontinuity across each first-entry cut, thereby
determining the corresponding Galois-odd second entry. Together, these steps
reconstruct the complete symbol directly from the dressed
representation.

\subsection{Leading singularities from the dressed representation}
\label{sec:dressedsymbol}
We now derive the symbol result~\eqref{eq:symres} from the following form of the dressed
representation:
\begin{equation}\label{eq:CdressedL}
C=\frac{x_1x_2x_3}{\pi^2}\int_{\mathbb R^2}
\frac{\,\dd q_1\wedge\dd q_2}
{(q_1^2{-}x_1^2)(q_2^2{-}x_2^2)(q_3^2{-}x_3^2)\sqrt{\Delta(q_1,q_2)}}\;\pT(q_1^2{-}y_1^2,q_2^2{-}y_2^2,q_3^2{-}y_3^2)\,.
\end{equation}
We remind the reader that $-q_3=q_1{+}q_2$, and that $\pT(P_1^2,P_2^2,P_3^2)$ is the pure-function part of the one-loop four-dimensional
triangle integral, {\it i.e.}, the function in square brackets of Eq.~\eqref{eq:TriUD}. In this calculation, we use its symbol. 
The root in the dressed representation is 
\begin{equation}\label{eq:DeltaDef}
\Delta(q_1,q_2) =\lambda(q_1^2-y_1^2,q_2^2-y_2^2,q_3^2-y_3^2)\,.
\end{equation}
Equation~\eqref{eq:CdressedL} is the Lorentzian form of the dressed representation introduced in
section~\ref{sec:dressed}. We analyse the singularities in this form, in which the auxiliary propagators read
$1/(p_i^2-x_i^2+i\varepsilon)$. Since the derivation below works at the level of the symbol,
overall constant normalisations are suppressed throughout this subsection.

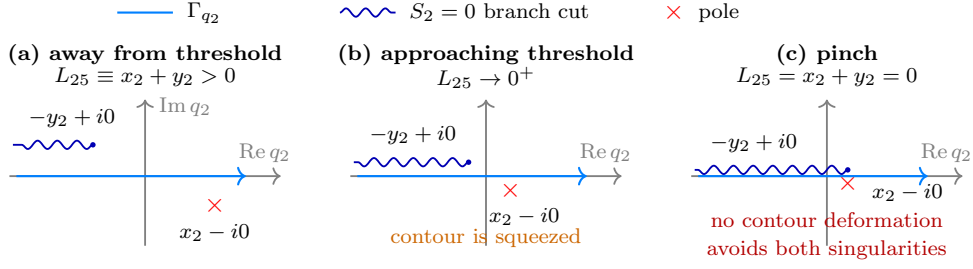
\begin{figure}
\centering
\begin{tikzpicture}[scale=0.92,line width=0.7pt,
  cut/.style={blue!70!black,decorate,decoration={snake,amplitude=0.55mm,segment length=2.6mm}},
  cont/.style={blue!50!cyan,thick},
  lbl/.style={font=\scriptsize},
  ttl/.style={font=\scriptsize\bfseries}]
\begin{scope}[shift={(-0.6,2.35)}]
\draw[cont] (0.2,0)--(1.0,0); \node[lbl,right] at (1.05,0) {$\Gamma_{q_2}$};
\draw[cut] (3.4,0)--(4.2,0); \node[lbl,right] at (4.25,0) {$S_2=0$ branch cut};
\node[red,font=\small] at (8.2,0) {$\times$}; \node[lbl,right] at (8.4,0) {pole};
\end{scope}
\begin{scope}[shift={(0,0)}]
\node[ttl] at (0,1.75) {(a) away from threshold};
\node[lbl] at (0,1.42) {$L_{25}\equiv x_2+y_2>0$};
\draw[->,gray] (-1.95,0)--(1.95,0);
\node[lbl,gray,above] at (1.72,0.05) {$\mathrm{Re}\,q_2$};
\draw[->,gray] (0,-1.0)--(0,1.1);
\node[lbl,gray,right] at (0.05,1.0) {$\mathrm{Im}\,q_2$};
\draw[cut] (-0.75,0.45)--(-1.9,0.45);
\fill[blue!70!black] (-0.75,0.45) circle (1.1pt);
\node[lbl,above] at (-1.05,0.55) {$-y_2+i0$};
\node[red,font=\small] at (1.0,-0.42) {$\times$};
\node[lbl,below] at (1.0,-0.54) {$x_2-i0$};
\draw[cont,->] (-1.85,0)--(1.45,0);
\end{scope}
\begin{scope}[shift={(4.9,0)}]
\node[ttl] at (0,1.75) {(b) approaching threshold};
\node[lbl] at (0,1.42) {$L_{25}\to0^+$};
\draw[->,gray] (-1.95,0)--(1.95,0);
\node[lbl,gray,above] at (1.72,0.05) {$\mathrm{Re}\,q_2$};
\draw[->,gray] (0,-1.0)--(0,1.1);
\draw[cut] (-0.25,0.2)--(-1.9,0.2);
\fill[blue!70!black] (-0.25,0.2) circle (1.1pt);
\node[lbl,above] at (-1.05,0.3) {$-y_2+i0$};
\node[red,font=\small] at (0.35,-0.2) {$\times$};
\node[lbl,below] at (0.55,-0.3) {$x_2-i0$};
\draw[cont,->] (-1.85,0)--(1.45,0);
\node[lbl,orange!80!black] at (0,-0.88) {contour is squeezed};
\end{scope}
\begin{scope}[shift={(9.8,0)}]
\node[ttl] at (0,1.75) {(c) pinch};
\node[lbl] at (0,1.42) {$L_{25}=x_2+y_2=0$};
\draw[->,gray] (-1.95,0)--(1.95,0);
\node[lbl,gray,above] at (1.72,0.05) {$\mathrm{Re}\,q_2$};
\draw[->,gray] (0,-1.0)--(0,1.1);
\draw[cut] (0.3,0.09)--(-1.9,0.09);
\fill[blue!70!black] (0.3,0.09) circle (1.1pt);
\node[lbl,above] at (-1.05,0.19) {$-y_2+i0$};
\node[red,font=\small] at (0.3,-0.1) {$\times$};
\node[lbl,right] at (0.5,-0.24) {$x_2-i0$};
\draw[cont,->] (-1.85,0)--(1.45,0);
\node[lbl,red!70!black,align=center] at (0,-0.85) {no contour deformation\\avoids both singularities};
\end{scope}
\end{tikzpicture}
\caption{Pinching of the $q_2$ contour and the locus $x_2+y_2\to0$. The pole and the branch
point approach the real contour from opposite sides; their collision produces the first-entry
divisor $x_2+y_2=0$.}
\label{fig:pinch}
\end{figure}

It is useful first to perform a partial-fraction decomposition of the prefactor
so that the integrand associated with each pure function $\Phi_i$ is separated. Constructing
the two-fold dlog forms yields
\begin{equation}\label{eq:dlogdecomp}
\frac{x_1x_2x_3\,\dd q_1\wedge\dd q_2}
{(q_1^2{-}x_1^2)(q_2^2{-}x_2^2)(q_3^2{-}x_3^2)\sqrt{\Delta(q_1,q_2)}}
=\sum_{i=1}^{6}R_i\ \dd\log(I_i)\wedge\dd\log(J_i)\,,
\end{equation}
Here, the $R_i$ are precisely the six leading singularities appearing in
the master formula~\eqref{eq:master}, while $I_i$ and $J_i$ are functions
of the auxiliary and external variables. Crucially, the $R_i$ depend only
on the external kinematics and can therefore be pulled outside the
integrations of $q_i$. Inserting~\eqref{eq:dlogdecomp} into
\eqref{eq:CdressedL}, and restoring the common normalisation, gives
\begin{equation}
C
=
\sum_{i=1}^{6} R_i
\left[
\frac{1}{\pi^2}
\int_{\mathbb R^2}
\dd\log(I_i)\wedge\dd\log(J_i)\,
\pT(q_1^2{-}y_1^2,q_2^2{-}y_2^2,q_3^2{-}y_3^2)
\right].
\end{equation}
The expression in square brackets is precisely the pure function
$\Phi_i$. Thus, the dlog decomposition~\eqref{eq:dlogdecomp} directly
reproduces the structure
\(
C=\sum_{i=1}^{6}R_i\Phi_i
\)
of the master formula~\eqref{eq:master} at the level of the dressed
integrand. In particular, it does more than identify the possible leading
singularities: it separates the full dressed representation into the six
sectors whose integrated coefficients are the six functions $\Phi_i$.

The explicit forms of $I_i$ and $J_i$ are rather cumbersome and will not
be needed below, so we do not display them. To illustrate how the symbol
of an individual sector can be extracted, we focus on the
coefficient of $R_1$. Expanding
$\dd\log(I_1)\wedge\dd\log(J_1)$ in terms of
$\dd q_1\wedge\dd q_2$, with the sign fixed by the chosen orientation,
gives
\begin{equation}\label{eq:Phi1int}
\;
\Phi_1=-\frac{1}{\pi^2}\int_{\mathbb R^2}
\frac{\sqrt{\Sigma_1}\,\big[x_2q_1+(x_2-x_3)q_2\big]\,
\dd q_1\wedge\dd q_2}
{4\,(q_2^2-x_2^2)\big((q_1+q_2)^2-x_3^2\big)
\sqrt{\Delta(q_1,q_2)}}\;
\pT(q_1^2{-}y_1^2,q_2^2{-}y_2^2,q_3^2{-}y_3^2)\,.
\;
\end{equation}
We now derive the symbol of $\Phi_1$ directly from that of
$\pT(q_1^2{-}y_1^2,q_2^2{-}y_2^2,q_3^2{-}y_3^2)$. The procedure has three steps: (1)
determine the pinches of the integral~\eqref{eq:Phi1int}, which give the first entries of
$\Phi_1$; (2) take the discontinuity on each cut and express it as a contour integral; and
(3) perform the remaining contour integral to obtain the odd letter following that first
entry.

\subsection{Determining all first entries}
\label{sec:firstentries}
A first-entry singularity occurs when the original integration contour can no longer be
deformed away from a set of singular divisors. This can be seen directly from~\eqref{eq:Phi1int}.

Consider first a pole--branch-cut pair in the $q_2$ plane. The two relevant loci in~\eqref{eq:Phi1int} are $D_2=q_2^2-x_2^2$, from the prefactor, and $S_2=q_2^2-y_2^2$,
from the first entries of the triangle. With the inherited prescription, the relevant local
singularities are the pole $q_2=x_2-i0$ and the branch point $q_2=-y_2+i0$. They approach the
real contour from opposite sides when $x_2+y_2\to0$, so the contour can no longer be deformed
between them and the integral develops a discontinuity (Fig.~\ref{fig:pinch}). The
conjugate pair gives the same locus. By contrast, the singularities that would lead to
$x_2-y_2=0$ remain on the same side of the contour on the physical sheet and do not pinch it.
Thus $x_2+y_2$, but not $x_2-y_2$, is a first entry of $\Phi_1$.

In total, the integrand gives the candidate divisors
\begin{equation}\label{eq:divisors}
D_2=q_2^2{-}x_2^2\,,\quad D_3=q_3^2{-}x_3^2\,,\quad
S_1=q_1^2{-}y_1^2\,,\quad S_2=q_2^2{-}y_2^2\,,\quad S_3=q_3^2{-}y_3^2\,.
\end{equation}
There are therefore six physical pinch circuits for $\Phi_1$:
\begin{table}
\centering
\caption{The six physical pinch circuits of the block $\Phi_1$ and their first-entry
letters.}
\label{tab:pinch}
\begin{tabular}{lll}
\toprule
pinching divisors & first-entry letter & interpretation\\
\midrule
$D_2,S_2$ & $x_2+y_2$ & one  pole and one triangle branch divisor\\
$D_3,S_3$ & $x_3+y_3$ & one  pole and one triangle branch divisor\\
$D_2,S_1,S_3$ & $x_2+y_1+y_3$ & one pole and two triangle branch divisors\\
$D_3,S_1,S_2$ & $x_3+y_1+y_2$ & one pole and two triangle branch divisors\\
$D_2,D_3,S_1$ & $x_2+x_3+y_1$ & two poles and one triangle branch divisor\\
$S_1,S_2,S_3$ & $y_1+y_2+y_3$ & three triangle branch divisors\\
\bottomrule
\end{tabular}
\end{table}
The fact that only sums occur on the physical sheet follows from the positive-energy
solutions selected by the common $i0$ prescription. Other sign choices describe crossed
sheets rather than additional first entries of the Euclidean correlator. These are exactly
the factors of the three first entries of $\Phi_1$, as shown in~\eqref{eq:Fconn1}.

Applying the same procedure to the dressed representation of each $\Phi_i$ yields the complete set of first entries in~\eqref{eq:Fconn1} and~\eqref{eq:Gfold4}.

\subsection{The discontinuity across the first entries}
\label{sec:discfirst}
Having determined the first entries, we can compute the discontinuity associated with each
of them directly from~\eqref{eq:Phi1int}. We again use $x_2+y_2$ as an example.

\begin{figure}
\centering
\begin{tikzpicture}[scale=0.92,line width=0.7pt,
  cut/.style={blue!70!black,decorate,decoration={snake,amplitude=0.55mm,segment length=2.6mm}},
  cont/.style={blue!50!cyan,thick},
  lbl/.style={font=\scriptsize},
  ttl/.style={font=\scriptsize\bfseries}]
\begin{scope}[shift={(-0.9,2.3)}]
\draw[cont] (0,0)--(0.8,0); \node[lbl,right] at (0.85,0) {integration contour};
\draw[cut] (4.0,0)--(4.8,0); \node[lbl,right] at (4.85,0) {$S_2=0$ branch cut};
\node[red,font=\small] at (8.4,0) {$\times$}; \node[lbl,right] at (8.6,0) {$q_2=x_2-i0$ pole};
\end{scope}
\begin{scope}[shift={(0,0)}]
\node[ttl] at (0,1.7) {original contour $\Gamma_{q_2}$};
\draw[->,gray] (-1.95,0)--(1.95,0);
\node[lbl,gray,above] at (1.72,0.05) {$\mathrm{Re}\,q_2$};
\draw[->,gray] (0,-1.0)--(0,1.15);
\node[lbl,gray,right] at (0.05,1.05) {$\mathrm{Im}\,q_2$};
\draw[cut] (-0.6,0.38)--(-1.9,0.38);
\fill[blue!70!black] (-0.6,0.38) circle (1.1pt);
\node[lbl,above] at (-0.95,0.48) {$-y_2+i0$};
\node[red,font=\small] at (0.7,-0.35) {$\times$};
\node[lbl,below] at (0.7,-0.48) {$x_2-i0$};
\draw[cont,->] (-1.85,0)--(1.45,0);
\node[lbl,align=center] at (0,-1.35) {pole and branch cut lie\\on opposite sides};
\end{scope}
\node at (2.4,0) {$=$};
\begin{scope}[shift={(4.9,0)}]
\node[ttl] at (0,1.7) {regular part $\Gamma_{\rm reg}$};
\draw[->,gray] (-1.95,0)--(1.95,0);
\node[lbl,gray,above] at (1.72,0.05) {$\mathrm{Re}\,q_2$};
\draw[->,gray] (0,-1.0)--(0,1.15);
\draw[cut] (-0.6,0.38)--(-1.9,0.38);
\fill[blue!70!black] (-0.6,0.38) circle (1.1pt);
\node[lbl,above] at (-0.95,0.48) {$-y_2+i0$};
\node[red,font=\small] at (0.7,-0.35) {$\times$};
\draw[cont,->] (-1.85,0)--(0.0,0) .. controls (0.35,0) and (0.35,-0.62) .. (0.7,-0.62)
 .. controls (1.05,-0.62) and (1.05,0) .. (1.35,0) -- (1.45,0);
\node[lbl,align=center] at (0,-1.35) {analytic in $x_2+y_2$};
\end{scope}
\node at (7.3,0) {$+$};
\begin{scope}[shift={(9.8,0)}]
\node[ttl] at (0,1.7) {localised part $\gamma_{x_2}$};
\draw[->,gray] (-1.95,0)--(1.95,0);
\node[lbl,gray,above] at (1.72,0.05) {$\mathrm{Re}\,q_2$};
\draw[->,gray] (0,-1.0)--(0,1.15);
\node[red,font=\small] at (0.7,-0.35) {$\times$};
\node[lbl,below] at (0.7,-0.75) {$x_2-i0$};
\draw[cont,->] (0.7,-0.35) ++(0:0.32) arc (0:330:0.32);
\node[lbl,align=center] at (0,-1.35) {contains only the $x_2$ pole};
\end{scope}
\end{tikzpicture}
\caption{Localisation of the $q_2$ contour:
$\Gamma_{q_2}=\Gamma_{\rm reg}+\gamma_{x_2}$, with the
integral over $\Gamma_{\rm reg}$ analytic in $x_2{+}y_2$.}
\label{fig:local}
\end{figure}
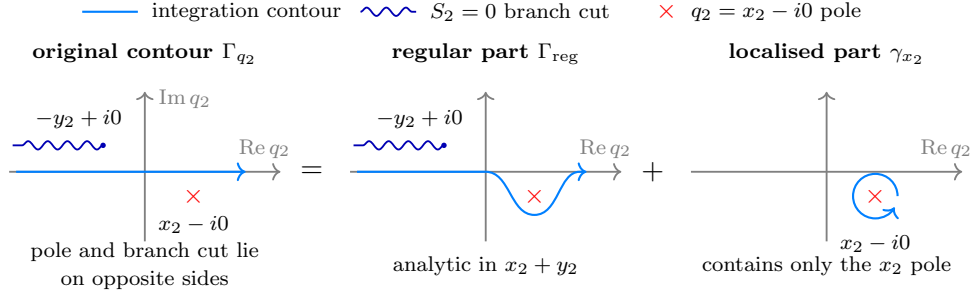

We first localise the $q_2$ contour.
The branch $x_2+y_2=0$ is generated by the pinch
between the pole $q_2=x_2-i0$ and the relevant branch point $q_2=-y_2+i0$. Isolating the discontinuity associated with this pinch, the full $q_2$
contour can be decomposed into two contours: an unbounded contour for which the pole and
branch cut remain on the same side, plus a small contour $\gamma_{x_2}$ surrounding only the
pole (Fig.~\ref{fig:local}). The first contour is analytic in $x_2+y_2$; hence only
$\gamma_{x_2}$ contributes to the discontinuity. To compute the discontinuity across $x_2+y_2$, we can therefore take the residue at
$q_2=x_2$ and retain only the remaining $q_1$ integral.

\begin{figure}
\centering
\begin{tikzpicture}[scale=0.92,line width=0.7pt,
  cut/.style={blue!70!black,decorate,decoration={snake,amplitude=0.55mm,segment length=2.6mm}},
  cont/.style={blue!50!cyan,thick},
  lbl/.style={font=\scriptsize},
  ttl/.style={font=\scriptsize\bfseries}]
\begin{scope}[shift={(-1.4,2.75)}]
\draw[cont] (0,0)--(0.8,0); \node[lbl,right] at (0.85,0) {integration contour};
\draw[cut] (4.0,0)--(4.8,0); \node[lbl,right] at (4.85,0) {$\log(\bar z/z)$ branch cut};
\node[red,font=\small] at (9.0,0) {$\times$}; \node[lbl,right] at (9.2,0) {excluded pole $q_0$};
\end{scope}
\begin{scope}[shift={(0,0)}]
\node[ttl] at (0,2.15) {the full $q_1$ contour $\Gamma_{q_1}$};
\draw[->,gray] (-2.6,0)--(2.75,0);
\node[lbl,gray,above] at (2.55,0.05) {$\mathrm{Re}\,q_1$};
\draw[->,gray] (0,-2.45)--(0,1.7);
\node[lbl,gray,left] at (-0.04,1.55) {$\mathrm{Im}\,q_1$};
\node[red,font=\small] at (-1.1,0.75) {$\times$};
\node[lbl,above] at (-1.1,0.87) {$q_0=-x_2-x_3+i0$};
\draw[cut] (-1.3,-0.28)--(1.3,-0.28);
\fill[blue!70!black] (-1.3,-0.28) circle (1.1pt);
\fill[blue!70!black] (1.3,-0.28) circle (1.1pt);
\node[lbl] at (-1.55,-0.55) {$q_C=-x_2+y_3-i0$};
\node[lbl] at (1.55,-0.55) {$q_A=y_1-i0$};
\draw[cont,->] (-2.3,0)--(2.3,0);
\draw[cont] (2.3,0) arc (0:-180:2.3);
\node[lbl,fill=white,inner sep=1pt] at (0,-1.9) {closing arc at infinity};
\end{scope}
\node at (3.55,0) {$=$};
\begin{scope}[shift={(7.9,0)}]
\node[ttl] at (0,2.15) {contour around the branch cut $\Gamma_{\text{branch}}$};
\draw[->,gray] (-2.6,0)--(2.75,0);
\node[lbl,gray,above] at (2.55,0.05) {$\mathrm{Re}\,q_1$};
\draw[->,gray] (0,-2.45)--(0,1.7);
\draw[cut] (-1.3,-0.28)--(1.3,-0.28);
\fill[blue!70!black] (-1.3,-0.28) circle (1.1pt);
\fill[blue!70!black] (1.3,-0.28) circle (1.1pt);
\node[lbl] at (-1.55,-0.75) {$q_C=-x_2+y_3-i0$};
\node[lbl] at (1.55,-0.75) {$q_A=y_1-i0$};
\draw[cont,->] (0,-0.28) ellipse (1.62 and 0.28);
\node[lbl,align=center] at (0,-1.55) {the contour is contracted\\onto the two lips of the cut};
\end{scope}
\end{tikzpicture}
\caption{Deformation of the $q_1$ contour: $\Gamma_{q_1}\simeq\Gamma_{\text{branch}}$, so the
integral of $\dd\log X\,\log(\bar z/z)$ can be taken along $\Gamma_{\text{branch}}$.}
\label{fig:deform}
\end{figure}
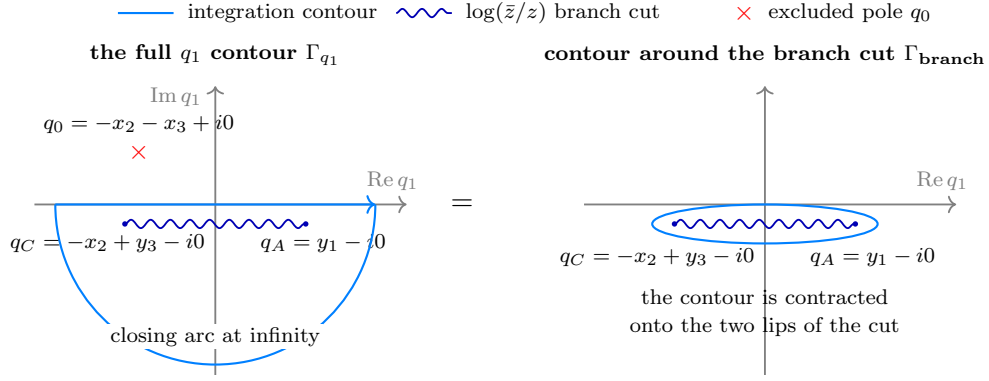

After localising the $q_2$ integration by taking the residue and then taking the discontinuity, we obtain
\begin{equation}\label{eq:disc1}
\mathrm{Disc}_{x_2+y_2}\Phi_1=
-\frac{\text{i}}{\pi}\int_{\mathbb R}\frac{\sqrt{\Sigma_1}\,\dd q_1}
{2(q_1+x_2+x_3)\sqrt{\Delta(q_1,x_2)}}\,
\log\Big(\frac{\bar z}{z}\Big)\bigg|_{q_2=x_2}\,.
\end{equation}
The remaining task is to perform this one-dimensional contour integral. The algebraic measure can be written in dlog form. With the definition below, one verifies
directly that
\begin{equation}\label{eq:dlogX}
-\frac{\sqrt{\Sigma_1}\,\dd q_1}{(q_1+x_2+x_3)\sqrt{\Delta(q_1,x_2)}}=\dd\log X(q_1)\,.
\end{equation}
where the $q_1$-dependent odd quantity is
\begin{equation}\label{eq:Xdef}
X(q_1)=\frac{\sqrt{\Sigma_1}+\sqrt{\Delta(q_1,x_2)}-2q_1y_2-2x_2y_2-2x_3y_2}
{\sqrt{\Sigma_1}-\sqrt{\Delta(q_1,x_2)}+2q_1y_2+2x_2y_2+2x_3y_2}\,.
\end{equation}
The triangle variables also satisfy
\begin{equation}\label{eq:zbarz}
\frac{\bar z}{z}\bigg|_{q_2=x_2}=
\frac{2q_1^2+2x_2q_1-y_1^2+y_2^2-y_3^2-\sqrt{\Delta(q_1,x_2)}}{2q_1^2+2x_2q_1-y_1^2+y_2^2-y_3^2+\sqrt{\Delta(q_1,x_2)}}\,,
\end{equation}
To perform the $q_1$ integration, note that, for $N_\pm=2q_1^2+2x_2q_1-y_1^2+y_2^2-y_3^2\pm\sqrt{\Delta(q_1,x_2)}$,
\begin{equation}\label{eq:PPDelta}
N_+\times N_-=4(q_1^2-y_1^2)((q_1+x_2)^2-y_3^2)\,,
\end{equation}
so the zeros and poles of $\bar z/z$, and hence the endpoints of its logarithmic cut, lie
over $q_1^2=y_1^2$ or $(q_1{+}x_2)^2=y_3^2$. The physical $i0$ prescription selects
\begin{equation}\label{eq:i0sel}
q_1=y_1-i0\,,\qquad q_1=-x_2+y_3-i0\,.
\end{equation}
The surviving pole $q_1=-x_2-x_3+i0$ is in the opposite half-plane. It is therefore outside
the deformation used for this sector. 
We can therefore perform the $q_1$ integration by closing the contour in
the lower half-plane, which contains the logarithmic cut. The resulting
contour can then be contracted onto the cut, so that the integral is given
by the integrals along its two sides with opposite orientations
(Fig.~\ref{fig:deform}), 
\begin{equation}\label{eq:lips}
\oint_{\Gamma_{\rm branch}}\dd\log X\,\log\frac{\bar z}{z}\bigg|_{q_2=x_2}
=\int_{\Gamma_{\rm upper}-\Gamma_{\rm lower}}\dd\log X\,\log\frac{\bar z}{z}\bigg|_{q_2=x_2}\,.
\end{equation}
However, across the two sides of its branch cut,
\begin{equation}\label{eq:jump2pi}
\log\frac{\bar z}{z}\bigg|_{+}-\log\frac{\bar z}{z}\bigg|_{-}=2\pi i\,.
\end{equation}
Therefore, the contour integral does not produce a new dilogarithm, and the result is
\begin{equation}\label{eq:endint}
\oint_{\Gamma_{\rm branch}}\dd\log X\,\log\frac{\bar z}{z}\bigg|_{q_2=x_2}
=2\pi i\int_{-x_2+y_3}^{y_1}\dd\log X
=2\pi i\,\log\frac{X(y_1)}{X(-x_2+y_3)}\,.
\end{equation}

To read off the letter on the right-hand side of~\eqref{eq:endint}, one must take care when evaluating the endpoint values. At each endpoint, the radicand $\Delta(q_1,x_2)$ becomes a perfect square:
\begin{equation}\label{eq:sqrtsub}
\begin{aligned}
\sqrt{\Delta(y_1,x_2)}&=\sqrt{\big(2x_2y_1+y_1^2+y_2^2-y_3^2\big)^2}\,,\\
\sqrt{\Delta(-x_2+y_3,x_2)}&=\sqrt{\big(2x_2y_3+y_1^2-y_2^2-y_3^2\big)^2}\,.
\end{aligned}
\end{equation}
The branches of
$\sqrt{\Delta(y_1,x_2)}$ and $\sqrt{\Delta(-x_2+y_3,x_2)}$ cannot be chosen by simply replacing
$\sqrt{Y^2}$ with $|Y|$. The square root is defined by analytic continuation on the same
sheet as the original contour. The cut must connect one zero and one pole of $\bar z/z$. From~\eqref{eq:zbarz}, the endpoint
at $q_1=y_1$ is chosen to obey $z=0$, whereas the endpoint at $q_1=-x_2+y_3$
obeys $\bar{z}=0$. This fixes
\begin{equation}\label{eq:sqrtfix}
\begin{aligned}
\sqrt{\Delta(y_1,x_2)}&=2x_2y_1+y_1^2+y_2^2-y_3^2\,,\\
\sqrt{\Delta(-x_2+y_3,x_2)}&=2x_2y_3+y_1^2-y_2^2-y_3^2\,.
\end{aligned}
\end{equation}
The opposite side of the cut reverses both signs simultaneously; it sends $\bar z/z$ and the resulting
odd letter each to its inverse, while also reversing the orientation. Substitution into~\eqref{eq:Xdef} gives
\begin{equation}\label{eq:Xratio}
\frac{X(y_1)}{X(-x_2+y_3)}=\frac{(y_1+y_3)^2-y_2^2+2x_3y_1+2(x_2+x_3)y_3-\sqrt{\Sigma_1}}{(y_1+y_3)^2-y_2^2+2x_3y_1+2(x_2+x_3)y_3+\sqrt{\Sigma_1}}
=\frac{1}{\chi^{(1)}_3}\,,
\end{equation}
Restoring the overall normalisation and the orientation
inherited from the original residue, one therefore obtains
\begin{equation}\label{eq:discres}
\mathrm{Disc}_{x_2+y_2}\Phi_1=-\log\chi^{(1)}_3\,.
\end{equation}
This is exactly the letter following the first entry $\log(x_2+y_2)$ obtained by expanding the
first entries in~\eqref{eq:symres}. Following a similar strategy, we can determine the
complete symbol of each $\Phi_i$, and therefore the symbol of the full correlator $C$, from the dressed representation alone.

Finally, passing from
the symbol \eqref{eq:symres} to the
function \eqref{eq:Phifun} is a standard step~\cite{Goncharov:2010jf}, and, as explained in
section~\ref{sec:compact}, the
beyond-the-symbol ambiguities are fixed by Galois parity and reality. This constitutes an
independent derivation of the closed form \eqref{eq:master}.

In summary, the dressed representation makes the symbol accessible without
performing the two auxiliary-energy integrations explicitly: its maximal
residues identify the six leading-singularity prefactors, the physical contour pinches determine the rational first entries, and the localised
discontinuities determine the corresponding Galois-odd second entries. The same procedure can, in principle, also be carried out directly in the cut
representation. There, the pole lines of the integration kernel, together
with the conic $\Omega(u,v)=0$ defined by the square root in the measure, form
a similar contour problem: extracting the discontinuity in each first-entry
channel again amounts to deforming the integration contours around the
relevant pinches. Thus, the dressed and cut representations expose the same symbol
data from complementary perspectives: contour discontinuities in the former
and projective incidence geometry in the latter.

\section{Bootstrapping the correlator from Landau singularities}
\label{sec:bootstrap}

The derivation in Section.~\ref{sec:bootprog}
determined the symbol directly from the dressed integral. In particular, it
determined the allowed pairings of first and last entries. In this section,
we show that the symbol can be determined even when less detailed information
is available. Once the leading singularities and the locations of the branch
points have been obtained from the dressed representation,
consistency conditions determine the symbol of the correlator uniquely. This type of symbol bootstrap based on Landau analysis has been studied in recent works on scattering amplitudes and Feynman integrals~\cite{Hannesdottir:2024hke,Chicherin:2025cua,Vergu:2025mag}.

The procedure is as follows. From the dressed representation, we determine a
set of rational first-entry letters $W_\alpha$, given by physical-sheet loci
formed from sums of the $x_i$ and $y_i$, as well as the square roots
$\sqrt{\Sigma_a}$ appearing in the leading singularities $R_a$. Instead of
obtaining the odd letters by explicitly computing every discontinuity of the
dressed representation, we search for polynomials $P_j^{(a)}$ in the $x_i$
and $y_i$ such that
\begin{equation}\label{eq:effortless}
 \bigl(P_j^{(a)}-\sqrt{\Sigma_a}\bigr)
 \bigl(P_j^{(a)}+\sqrt{\Sigma_a}\bigr)
 =
 c_j^{(a)}
 \prod_\alpha Y_\alpha^{\,n_{\alpha j}^{(a)}} ,
\end{equation}
where $c_j^{(a)}\in\mathbb{Q}^{\times}$ and the rational factors $Y_\alpha$
are drawn from the candidate Landau loci
\begin{equation}\label{eq:effortlessloci}
 \bigl\{
 x_i\pm y_i,\quad
 x_i\pm y_j\pm y_k,\quad
 x_i\pm x_j\pm y_k,\quad
 y_1\pm y_2\pm y_3
 \bigr\},
 \qquad i,j,k\ \text{distinct}.
\end{equation}
Although only the all-plus combinations occur as first entries on the
physical sheet, the signed loci are required to close the rational norm
$\bigl(P_j^{(a)}\bigr)^2-\Sigma_a$ under factorisation and include crossed or
secondary-sheet Landau divisors. This search is implemented, for example, in the package \texttt{Effortless}~\cite{Effortless}.
It constructs candidate odd letters of the form
\begin{equation}
\frac{P_j^{(a)}+\sqrt{\Sigma_a}}
      {P_j^{(a)}-\sqrt{\Sigma_a}} .
\end{equation}
For each of the roots $\sqrt{\Sigma_a}$ with $a=1,2,3$, the search yields
three independent odd letters, giving the nine letters
$\chi_j^{(a)}$ displayed in~\eqref{eq:chi1} and their symmetry images. For
each of the roots with $a=4,5,6$, it similarly yields four independent odd
letters, giving the twelve letters $\upsilon_j^{(a)}$.

We can now construct the symbol of each $\Phi_a$ by imposing integrable condition.
A general weight-$w$ tensor is the symbol of an analytic function if and only if it satisfies
\begin{equation}
\begin{aligned}
\sum_{i_1,\ldots,i_w} c_{i_1\cdots i_w}&\times \bigl(\dd\log \mathcal{L}_{i_k}\wedge
      \dd\log \mathcal{L}_{i_{k+1}}\bigr)\times\,\\
&\mathcal{L}_{i_1}\otimes\cdots\otimes \mathcal{L}_{i_{k-1}}
\otimes
\mathcal{L}_{i_{k+2}}\otimes\cdots\otimes \mathcal{L}_{i_w}
=0,
\qquad k=1,\ldots,w-1 .
\end{aligned}
\label{eq:integrability}
\end{equation}
Since each $\Phi_a$ is a weight-two function whose symbol is odd under the
Galois involution associated with $\sqrt{\Sigma_a}$, we make the ansatz
\begin{equation}
 \cS(\Phi_a)
 =
 \sum_{\alpha,j}c_{\alpha j}^{(a)}
 W_\alpha\otimes\chi_j^{(a)}
\end{equation}
for $a=1,2,3$, and the analogous ansatz with
$\chi_j^{(a)}$ replaced by $\upsilon_j^{(a)}$ for $a=4,5,6$. Here, $c_{\alpha j}^{(a)}$ is a matrix of rational coefficients to be determined. Integrable condition reduces the bootstrap to the linear problem of finding all allowed matrices $c_{\alpha j}^{(a)}$ satisfying
\begin{equation}
 \sum_{\alpha,j}c_{\alpha j}^{(a)}
 \dd\log W_\alpha\wedge\dd\log\chi_j^{(a)}=0 .
\end{equation}
A calculation shows that, once odd letters associated with a
single square root $\sqrt{\Sigma_a}$ are used, the solution space is one-dimensional for each $a$. For instance, for the case $a=1$, we take as candidate first entries all
ten all-plus factors contained in the general set~\eqref{eq:effortlessloci}, Together with the three independent odd letters $\chi_j^{(1)}$, this
gives a $10\times3$ ansatz containing thirty rational coefficients.
Solving the integrable condition leaves
a one-dimensional solution space and therefore fixes the symbol uniquely
up to an overall normalisation.
This determines the symbol of each
$\Phi_a$. The corresponding function-level expressions still require the
beyond-the-symbol information, which is fixed here by Galois parity,
reality, and the branch prescription, as discussed in
section~\ref{sec:compact}.

Finally, to reconstruct the full correlator $C$ in~\eqref{eq:master}, we
impose its symmetries. The correlator is a sum of the six leading
singularities $R_a$ multiplying the real functions $\Phi_a$. It is invariant
under the $S_3$ permutations of the vertices, which fixes the relative
coefficients among the three terms within each of the two orbits. The conformal Ward identities
\eqref{eq:euler}--\eqref{eq:cwipair} then fix the relative coefficient
between the two orbits. The remaining overall normalisation can be fixed by
one boundary datum, for instance the total-energy residue
\eqref{eq:residue}. These conditions are in fact overdetermined: the Ward
identities act letter by letter and are satisfied identically by the
surviving combination, providing strong consistency relations beyond those
needed to fix the answer. 

This turns the present computation into a potential strategy for higher-point
one-loop correlators, for which the dressing integrals are
higher-dimensional and remain difficult to evaluate directly, even after
taking a discontinuity to reduce the number of integrations. On the other hand, similar inputs for their Landau bootstrap are within reach. For example, for the one-loop four-site correlator, the
leading singularities and algebraic units can be obtained from the dressed
representation, the branch-point loci from Landau analysis, and candidate
odd letters from the algorithm in~\eqref{eq:effortless}. The full correlator
must also satisfy the appropriate permutation symmetries and conformal Ward
identities. A more basic question, however, must first be addressed: whether
the box can still be expressed in terms of multiple polylogarithms or
requires more general transcendental functions, what its transcendental
weight is, and whether it has uniform weight. We expect that these questions
can also be explored using the dressed representation. A detailed analysis
is beyond the scope of the present paper and is left for future work.

\acknowledgments
This project has received funding from the European Research Council (ERC) under the European
Union's Horizon Europe research and innovation programme (grant agreement No.\ 101097219,
UNIVERSE\raisebox{0.2ex}{+}). Views and opinions expressed are however those of the authors only and
do not necessarily reflect those of the European Union or the European Research Council Executive
Agency. Neither the European Union nor the granting authority can be held responsible for them. We
also acknowledge support from the Max Planck-IAS-NTU Center for Particle Physics, Cosmology and
Geometry.

\appendix

\section{Stieltjes and complete-monotonicity properties}
\label{app:positivity}

In this appendix, we make precise several positivity properties of the
triangle correlator. Similar properties of Euclidean Feynman integrals
have recently been studied in
refs.~\cite{Henn:2024qwe,Ditsch:2025rdx,Raman:2026ect}.
It is convenient to remove the elementary kinematic prefactor and define
\begin{equation}
    \widehat C
    \equiv
    \frac{C}{x_1x_2x_3}.
\end{equation}
We will show that $\widehat C$ is a Stieltjes function of each individual
squared energy, with all other variables held fixed, and that it is
jointly completely monotone in the six variables
\begin{equation}
    \boldsymbol{s}
    =
    \bigl(
    x_1^2,x_2^2,x_3^2,
    y_1^2,y_2^2,y_3^2
    \bigr).
\end{equation}
Although the physical variables $y_i$ obey triangle inequalities, the
integral representation below defines a positive analytic extension in
which the six squared energies can be varied independently over the
positive orthant. The statements in this appendix refer to this
extension and therefore hold, in particular, in the physical Euclidean
region.

A Stieltjes function of one variable admits a representation of the form
\begin{equation}
    f(s)
    =
    \alpha
    +
    \int_0^\infty
    \frac{\mathrm{d}\rho(t)}{s+t},
    \qquad
    \alpha\geq 0,
    \qquad
    \mathrm{d}\rho(t)\geq 0.
\end{equation}
A function of several positive variables is jointly completely monotone
if
\begin{equation}
    (-1)^{|\boldsymbol{n}|}
    \frac{\partial^{|\boldsymbol{n}|}f}
    {\partial s_1^{n_1}\cdots\partial s_6^{n_6}}
    \geq 0
    \qquad
    \text{for all }
    \boldsymbol{n}\in\mathbb{N}_0^6,
\end{equation}
where $|\boldsymbol{n}|=\sum_{r=1}^6 n_r$.

\paragraph{Stieltjes property in each variable.}

Inserting the Feynman-parameter representation of the flat-space
triangle~\eqref{eq:Tri} into the dressed
representation~\eqref{eq:Cdressed}, we obtain
\begin{equation}
\begin{split}
    \widehat C
    &=
    \frac{1}{\pi^2}
    \int_{\mathbb{R}^2}
    \mathrm{d}q_1\,\mathrm{d}q_2
    \int_{a_i\geq 0}
    \mathrm{d}^3a\,
    \delta\!\left(1-\sum_{i}a_i\right)
   \times
    \frac{1}{
    (q_1^2+x_1^2)
    (q_2^2+x_2^2)
    (q_3^2+x_3^2)}
    \frac{1}{\mathcal{F}},
\end{split}
\label{eq:sixfold}
\end{equation}
where
\begin{equation}
    \mathcal{F}
    =
    a_1a_2(q_1^2+y_1^2)
    +
    a_2a_3(q_2^2+y_2^2)
    +
    a_3a_1(q_3^2+y_3^2).
\label{eq:Fpoly}
\end{equation}
The integrand is non-negative throughout the Euclidean integration
region.

Consider first one of the variables $s=x_i^2$. Its entire dependence is
contained in a single propagator,
\begin{equation}
    \frac{1}{q_i^2+s},
\end{equation}
while all the remaining factors are non-negative and independent of
$s$. Integration over the remaining variables therefore produces a
positive superposition of Stieltjes kernels. The same argument applies to the variables $s=y_i^2$. For fixed
Feynman parameters and auxiliary energies, one may write
\begin{equation}
    \mathcal{F}
    =
    \mathcal{F}_{(i)}+c_i s,
    \qquad
    (c_1,c_2,c_3)
    =
    (a_1a_2,a_2a_3,a_3a_1),
\end{equation}
where $\mathcal{F}_{(i)}$ is independent of $s$ and non-negative. In the
interior of the Feynman-parameter simplex, $c_i>0$, and hence
\begin{equation}
    \frac{1}{\mathcal{F}_{(i)}+c_i s}
    =
    \frac{1}{c_i}
    \frac{1}{s+\mathcal{F}_{(i)}/c_i}.
\end{equation}
This is again a Stieltjes kernel with a non-negative coefficient. The
boundaries on which $c_i=0$ are obtained by continuity and do not alter
the integral. We conclude that $\widehat C$ is an ordinary Stieltjes
function of each of the six variables $x_i^2$ and $y_i^2$ separately,
when the other five variables are held fixed.

\paragraph{A multivariate Stieltjes representation.}

The joint dependence on all six squared energies is most transparent in
a fully parametric representation. Introducing Schwinger parameters for
the three dressing propagators and carrying out the two Gaussian
auxiliary-energy integrations gives
\begin{equation}
\begin{split}
    C
    &=
    \frac{2x_1x_2x_3}{\pi}
    \int_0^\infty
    \mathrm{d}^3a\,\mathrm{d}^3b\,
    \delta\!\left(
    1-\sum_ia_i-\sum_ib_i
    \right)
    \frac{A^{3/2}}{\sqrt{B}\,\mathcal{F}^3},
\end{split}
\label{eq:paramFeynman}
\end{equation}
where
\begin{equation}
    A=a_1+a_2+a_3,
\label{eq:paramA}
\end{equation}
\begin{equation}
\begin{split}
    B={}&
    a_1a_2a_3
    +a_1a_2(b_2+b_3)
    +a_2a_3(b_1+b_3)
    +a_3a_1(b_1+b_2)
    \\
    &\quad
    +A(b_1b_2+b_2b_3+b_3b_1),
\end{split}
\label{eq:paramB}
\end{equation}
and
\begin{equation}
\begin{split}
    \mathcal{F}
    ={}&
    a_1a_2y_1^2
    +a_2a_3y_2^2
    +a_3a_1y_3^2  
    +A\left(
    b_1x_1^2+b_2x_2^2+b_3x_3^2
    \right).
\end{split}
\label{eq:paramF}
\end{equation}
All coefficients appearing in $\mathcal{F}$ are non-negative on the
integration domain. Defining
\begin{equation}
    \boldsymbol{c}(a,b)
    =
    \bigl(
    Ab_1,Ab_2,Ab_3,
    a_1a_2,a_2a_3,a_3a_1
    \bigr),
\end{equation}
we can write
\begin{equation}
    \mathcal{F}
    =
    \boldsymbol{c}(a,b)\cdot\boldsymbol{s}.
\end{equation}
Consequently,
\begin{equation}
    \widehat C(\boldsymbol{s})
    =
    \int
    \frac{\mathrm{d}\mu(a,b)}
    {\bigl(\boldsymbol{c}(a,b)\cdot\boldsymbol{s}\bigr)^3},
\label{eq:multivariateStieltjes}
\end{equation}
with the non-negative measure
\begin{equation}
\begin{split}
    \mathrm{d}\mu(a,b)
    ={}&
    \frac{2}{\pi}\,
    \mathrm{d}^3a\,\mathrm{d}^3b\,
    \delta\!\left(
    1-\sum_ia_i-\sum_ib_i
    \right)
    \frac{A^{3/2}}{\sqrt{B}}.
\end{split}
\end{equation}
Equation~\eqref{eq:multivariateStieltjes} is a homogeneous generalized
multivariate Stieltjes representation of order three: it expresses
$\widehat C$ as a positive superposition of the kernels
$(\boldsymbol{c}\cdot\boldsymbol{s})^{-3}$.

The joint complete monotonicity now follows directly. For any
$\boldsymbol{n}\in\mathbb{N}_0^6$,
\begin{equation}
\begin{split}
    (-1)^{|\boldsymbol{n}|}
    \frac{
    \partial^{|\boldsymbol{n}|}\widehat C
    }{
    \partial s_1^{n_1}\cdots\partial s_6^{n_6}
    }
    &=\frac{\Gamma(3+|\boldsymbol{n}|)}{\Gamma(3)}
    \int
    \mathrm{d}\mu(a,b)\,
    \frac{
    \displaystyle\prod_{r=1}^6
    c_r(a,b)^{n_r}
    }{
    \bigl(
    \boldsymbol{c}(a,b)\cdot\boldsymbol{s}
    \bigr)^{3+|\boldsymbol{n}|}
    }
    \geq 0.
\end{split}
\label{eq:jointCompleteMonotonicity}
\end{equation}
Differentiation under the integral is
valid in the interior of the positive Euclidean region.

Thus, the mixed representation~\eqref{eq:sixfold} shows that
$\widehat C$ is an ordinary Stieltjes function of each squared energy
separately, while the fully parametric
representation~\eqref{eq:multivariateStieltjes} identifies it as a
generalized multivariate Stieltjes transform of order three and proves
its joint complete monotonicity.

\bibliographystyle{utphys} 
\bibliography{ref.bib} 

\end{document}